\documentclass[aps,prc,preprint]{revtex4}

\usepackage{graphicx}% Include figure files
\usepackage{dcolumn}% Align table columns on decimal point
\usepackage{bm}% bold math

\begin{document}

%\preprint{APS/123-QED}

\title{Isoscalar giant resonances in the Sn nuclei and implications for the 
asymmetry term in the nuclear-matter incompressibility}

\author{T. Li}
\affiliation{Physics Department, University of Notre
Dame, Notre Dame, IN 46556, USA}
\author{U. Garg} 
\affiliation{Physics Department, University of Notre
Dame, Notre Dame, IN 46556, USA}
\author{Y. Liu} 
\affiliation{Physics Department, University of Notre
Dame, Notre Dame, IN 46556, USA}
\author{R. Marks}
\affiliation{Physics Department, University of Notre Dame, Notre Dame, IN 46556, USA}
\author{B.K. Nayak}
\affiliation{Physics Department, University of Notre Dame, Notre Dame, IN 46556, USA}
\author{P. V. Madhusudhana Rao}
\affiliation{Physics Department, University of Notre Dame, Notre Dame, IN 46556, USA}
\author{M. Fujiwara}
\affiliation{Research Center for Nuclear Physics, Osaka University, Mihogaoka Ibaraki 10-1 Osaka, 567-0047 Japan}
\author{H. Hashimoto}
\affiliation{Research Center for Nuclear Physics, Osaka University, Mihogaoka Ibaraki 10-1 Osaka, 567-0047 Japan}
\author{K. Nakanishi}
\affiliation{Research Center for Nuclear Physics, Osaka University, Mihogaoka Ibaraki 10-1 Osaka, 567-0047 Japan}
\author{S. Okumura} 
\affiliation{Research Center for Nuclear Physics, Osaka University, Mihogaoka Ibaraki 10-1 Osaka, 567-0047 Japan}
\author{M. Yosoi}
\affiliation{Research Center for Nuclear Physics, Osaka University, Mihogaoka Ibaraki 10-1 Osaka, 567-0047 Japan}
\author{M. Ichikawa}
\affiliation{Cyclotron and Radioisotope Center, Tohoku University, Sendai 980-8578, Japan}
\author{M. Itoh} 
\affiliation{Cyclotron and Radioisotope Center, Tohoku University, Sendai 980-8578, Japan}
\author{R. Matsuo}
\affiliation{Cyclotron and Radioisotope Center, Tohoku University, Sendai 980-8578, Japan}
\author{T. Terazono}
\affiliation{Cyclotron and Radioisotope Center, Tohoku University, Sendai 980-8578, Japan}
\author{M. Uchida}
\affiliation{Tokyo Institute of Technology, 2-12-1 O-Okayama, Tokyo 152-8550, Japan}
\author{Y. Iwao}
\affiliation{Department of Physics, Kyoto University, Kyoto 606-8502, Japan}
\author{T. Kawabata}
\affiliation{Department of Physics, Kyoto University, Kyoto 606-8502, Japan}
\author{T. Murakami}
\affiliation{Department of Physics, Kyoto University, Kyoto 606-8502, Japan}
\author{H. Sakaguchi}
\affiliation{Department of Physics, Kyoto University, Kyoto 606-8502, Japan}
\author{S.~Terashima}
\affiliation{Department of Physics, Kyoto University, Kyoto 606-8502, Japan}
\author{Y. Yasuda}
\affiliation{Department of Physics, Kyoto University, Kyoto 606-8502, Japan}
\author{J. Zenihiro}
\affiliation{Department of Physics, Kyoto University, Kyoto 606-8502, Japan}
\author{H. Akimune}
\affiliation{Department of Physics, Konan University, Hyogo, 658-8501 Japan}
\author{K. Kawase}
\affiliation{Advanced Photon Research Center, Japan Atomic Energy Agency,
Kizugawa, Kyoto 619-0215, Japan}
\author{M. N. Harakeh}
\affiliation{Kernfysisch Versneller Instituut, University of Groningen, 
9747 AA Groningen, The Netherlands}
\affiliation{GANIL, CEA/DSM - CNRS/IN2P3, F-14076 Caen, France}

\begin{abstract}

We have investigated the isoscalar giant resonances in the Sn
isotopes using inelastic scattering of 386-MeV $\alpha$-particles
at extremely forward angles, including 0$^\circ$. We have obtained completely
``background-free'' inelastic-scattering spectra for the Sn isotopes
over the angular range $0^\circ$-$9^\circ$ and up to an excitation
energy of 31.5 MeV. The strength distributions for various
multipoles were extracted by a multipole decomposition analysis
based on the expected angular distributions of the respective
multipoles. We find that
the centroid energies of the isoscalar giant monopole resonance (ISGMR)
in the Sn isotopes are significantly lower than the theoretical
predictions.
In addition, based on the ISGMR results, a value of $K_{\tau} = -550 \pm 100$ MeV is obtained for
the asymmetry term in the nuclear incompressibility. Constraints on interactions employed in nuclear structure calculations are discussed on the basis of the experimentally-obtained values for
$K_{\infty}$ and $K_{\tau}$.

\end{abstract}


\date{\today}

\maketitle

\section{Introduction}

Isoscalar giant resonances have been extensively studied since the
discovery of the isoscalar giant quadrupole resonance (ISGQR) in the early 1970s
\cite{rp1,futo,lebe}. The isoscalar giant monopole resonance (ISGMR) was
identified in 1977 \cite{mnh2,dhy2} and was the subject of a number of studies
through the 1980s \cite{av1,dhy1,sharma}. The isoscalar giant dipole resonance (ISGDR) was
first reported by Morsch {\it {\em et al.}} \cite{hpm1} in $^{208}$Pb but
was conclusively identified by Davis {\it {\em et al.}} \cite{bfd1}. Both ISGMR and ISGDR
are classified as compression modes and provide information about nuclear
incompressibility, $K_{A}$, from which the incompressibility of infinite nuclear
matter, K$_\infty$, may be obtained \cite{th-JPB-1}. 

Most of the earlier investigations of the isoscalar giant resonances used inelastic 
$\alpha$ scattering at 100--200 MeV and
the strength of a particular giant resonance was assumed
to be concentrated in a single peak with a Gaussian or Lorentzian shape. The
resonance parameters were obtained by multiple-peak fits to the inelastic scattering spectra, after
subtraction of a suitable ``background'' ~\cite{speth,mnh1}.
In the last decade, the Texas A\&M (TAMU) group has carried out
($\alpha, \alpha'$) studies of many nuclei at a bombarding energy of
240 MeV and extracted the strength distributions of various isoscalar 
giant resonances in a number of nuclei
\cite{exp-TAM-8,exp-TAM-9,exp-TAM-7,exp-TAM-1,
john,exp-TAM-3,exp-TAM-2,exp-TAM-6,exp-TAM-5,exp-TAM-11,exp-TAM-10}
using a multipole decomposition analysis (MDA)~\cite{exp-bb-1}.
Contemporaneously,
we have carried out giant resonance measurements using inelastic scattering of 386 MeV
$\alpha$ particles at extremely small angles, including 0$^{\circ}$ 
\cite{ugberk,exp-RCNP-3,exp-RCNP-2,exp-RCNP-5,exp-RCNP-1,exp-ND-3,itoh,exp-ND-2,tao,ug7}. 
An especially useful feature of our
measurements has been the elimination of all instrumental background events 
from the inelastic scattering
spectra which was rendered possible by the optical properties of the Grand
Raiden spectrometer~\cite{exp-RCNP-6}.

Here, we report on measurements of the isoscalar
giant resonances in the even-A Sn isotopes (A=112--124)
over the excitation-energy range 8.5--31.5 MeV.  Previously, giant resonance measurements on the Sn isotopes have been reported by the TAMU group \cite{dhy1,dhy4} and KVI group \cite{sharma},
using inelastic $\alpha$ scattering  at 120--130 MeV and peak-fitting analyses of spectra. More recently, the strength distributions of various isoscalar resonances have been obtained in some Sn isotopes
(A=112,116,124) by the TAMU group \cite{exp-TAM-3,exp-TAM-6}. 

High-quality measurements of the 
ISGMR over the full range of isotopes provide the opportunity to investigate the asymmetry term, 
$K_{\tau}$, of the nuclear
incompressibility. This term, associated with the neutron excess ($N - Z$), is crucial in obtaining the
radii of neutron stars in equation-of-state (EOS) calculations~\cite{exp-JML-1,latti,stein,exp-BAL-2}; the asymmetry ratio, [$(N - Z)/A$], changes by more than 80\% over the range of the investigated Sn isotopes.

\section{Experimental Techniques}

The experiment was performed at the ring cyclotron facility of the
Research Center for Nuclear Physics (RCNP), Osaka University, using
inelastic scattering of 386-MeV $\alpha$ particles at extremely
forward angles, including 0$^\circ$. 
A $^4$He$^{++}$ beam was accelerated by the Azimuthally
Varying Field (AVF) cyclotron up to 86 MeV, injected into the K = 400
ring cyclotron for acceleration up to 386 MeV, and achromatically transported to the WS
experimental hall without any defining slits. To reduce the background at
and near 0$^{\circ}$, the beam halo has to be tuned carefully in the
experiment. This was accomplished by tuning the beam profile of the
injection beam for the ring cyclotron, and typically less than 1 out of 10000
events had contamination from the other bunches. The beam current
was 1-20 nA, which was limited by the data acquisition rate and by
the maximum available current of the accelerator. The energy
resolution obtained was $\sim150$ keV full width at half maximum (FWHM).

Self supporting target foils of enriched even-A $^{112-124}$Sn isotopes of thickness
5.0--9.25 mg/cm$^2$ were employed; we used special target frames with a large
aperture in order to
reduce the background caused by the beam-halo hitting the frames. 
Data were also taken with a $^{nat}$C target at the actual field
settings used in the experiments and the energy calibration was
obtained from the peak positions of the 7.652- and 9.641-MeV states in 
the $^{12}$C$(\alpha,\alpha')$ spectra.

Inelastically-scattered
$\alpha$ particles were momentum-analyzed with the high-resolution
magnetic spectrometer, ``Grand Raiden'' \cite{exp-RCNP-6}, and the
vertical and horizontal positions of the $\alpha$ particles were
measured with the focal-plane detector system comprised of two
position-sensitive multi-wire drift chambers (MWDCs) and two scintillators \cite{exp-RCNP-5}.
The MWDCs and scintillators enabled us to make particle identification and to reconstruct 
the trajectories of the scattered particles. The scattering angle at the target and the
momentum of the scattered particles were determined by the
ray-tracing method. The vertical-position spectrum obtained in the
double-focusing mode of the spectrometer was exploited to eliminate
the instrumental background \cite{exp-RCNP-2, exp-RCNP-5}.

Giant-resonance data were taken with the spectrometer central angle
($\theta_{spec}$) set at 0$^\circ$, 2.5$^\circ$, 3.5$^\circ$,
5.0$^\circ$, 6.5$^\circ$, and 8.0$^\circ$, covering the angular range
from 0$^\circ$ to 9.0$^\circ$ in the laboratory system. The actual angular 
resolution of the MWDCs,
including the nominal broadening of scattering angle due to the emittance of
the $^4$He$^{++}$ beam and the multiple Coulomb-scattering effects, was
about 0.15$^\circ$ \cite{itohthesis}. The vertical acceptance was limited to $\pm$20
mr by a 2-mm-thick tantalum collimator. The energy bite of the Grand
Raiden spectrometer and the special MWDC arrangements for the 0$^{\circ}$ measurements
limited the excitation energy range observed to $E_x$=8.5--31.5 MeV. 

The incident $^4{\rm He}^{++}$ beam was stopped at three independent Faraday Cups (FC)
according to the different settings of Grand Raiden \cite{exp-RCNP-5}. In the
measurements at large angles ($\theta \geq 6.5^{\circ}$), the beam
was stopped on the FC mounted inside the scattering
chamber (SC-FC). 
For measurements at $0^{\circ}$, the FC was located downstream of the
MWDCs \cite{exp-RCNP-5}; the incident $^4$He$^{++}$ particles were guided to a vacuum pipe situated at the high-momentum side of the MWDCs and finally stopped at the $0^{\circ}$-FC. A third
FC was used for measurements in the scattering-angle region
$2^{\circ} \leq \theta \leq 5^{\circ}$.
This FC was
installed behind the Q1-magnet of the Grand
Raiden (Q1-FC)~\cite{itoh-rep}. The use of these three Faraday-cups
allowed us to obtain reliable values of accumulated charges for the incident
$^4{\rm He}^{++}$ beam at different scattering angles. Normalization of the
FCs was obtained with a beamline polarimeter located upstream from the target.
The polarimeter target was inserted in the beam, and the scattered
$^{4}$He$^{++}$ particles counted, before and after each change of the FC.
The overall accuracy of this normalization is estimated to be $\sim$2\%,
including systematic errors from slight changes in the direction of the beam
during the measurement \cite{itohthesis}.

\begin{figure}[htpb]
\begin{center}
\includegraphics[scale=0.8]{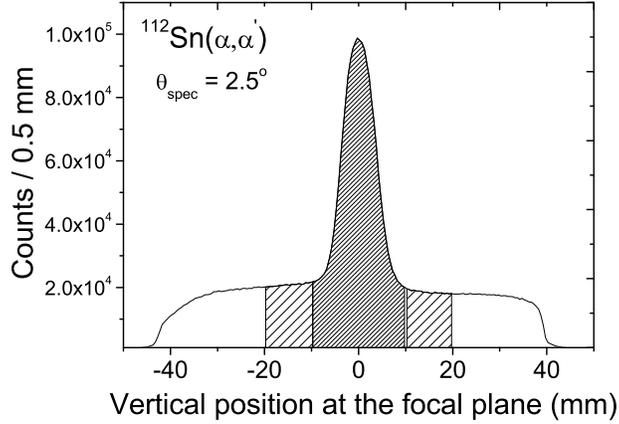}
\caption{\label{fig:ypos} Vertical-position spectrum at the focal
plane of the Grand Raiden spectrometer, taken at 2.5$^\circ$. The central densely-hatched region
represents true+background events. The off-center sparsely-hatched regions represent only
background events. The real events were obtained by subtracting
background events from the true+background events.}
\end{center}
\end{figure}

\begin{figure}[htpb]
\begin{center}
\includegraphics[scale=0.8]{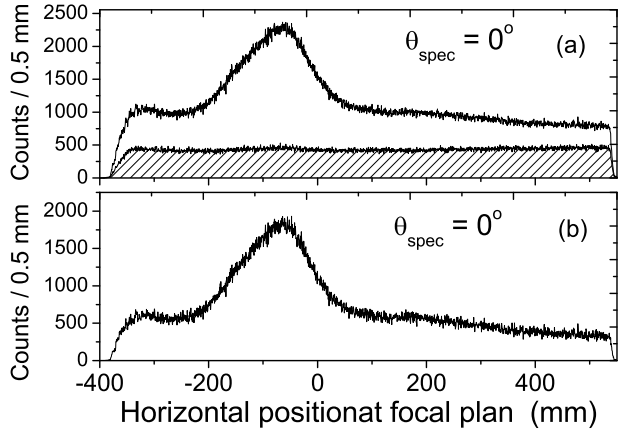}
\caption{\label{fig:xpos} (a) Horizontal-position spectrum of
the $^{112}$Sn($\alpha$,$\alpha'$) reaction at 0$^\circ$. The hatched
region is background events. (b) Background-free spectrum }
\end{center}
\end{figure}

The ion-optics of the Grand Raiden spectrometer is such that the particles scattered from the
target position are focused vertically and horizontally at the
focal plane. Using this property, the instrumental background was
completely eliminated. While inelastically scattered $\alpha$
particles are focused at the focal plane, background events due to
the rescattering of $\alpha$ particles from the wall and pole
surfaces of the spectrometer show a flat distribution in the
vertical-position spectra at the focal plane, as shown in
Fig.~\ref{fig:ypos}. The peak region in the vertical position spectrum was treated as
true+background events. The off-center regions were treated as
background only. Figure~\ref{fig:xpos}(a) shows the horizontal
position spectrum for the $^{112}$Sn($\alpha,\alpha'$) reaction at
0$^{\circ}$. The background spectrum has no distinct structure in
the giant resonance region. Finally, we have obtained clean spectra
by subtracting the background spectrum from the true+background
spectrum, as shown in Fig.~\ref{fig:xpos}(b).

\begin{figure}[htpb]
\begin{center}
\includegraphics[scale=1.1]{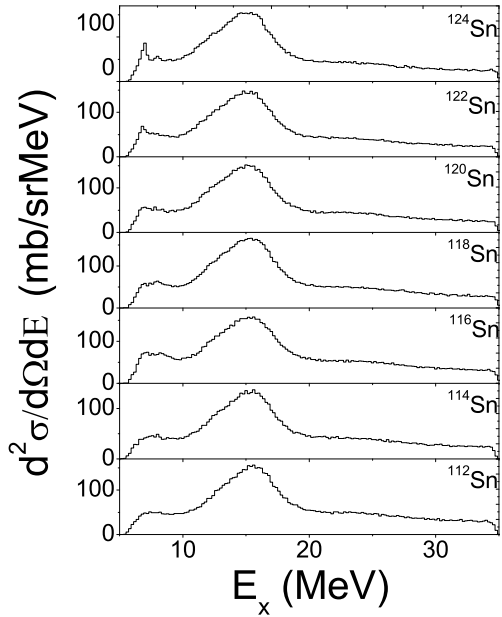}
\caption{\label{fig:sn0deg} Excitation-energy spectra obtained from inelastic $\alpha$ scattering at
  $\theta_{lab}$ = 0.69$^\circ$ for all even-A Sn isotopes.}
\end{center}
\end{figure}

The background-free ``$0^\circ$'' inelastic spectra for the Sn
isotopes are presented in Fig.~\ref{fig:sn0deg}. In all cases, the
spectrum is dominated by the ISGMR+ISGQR peak near $E_x\sim15$ MeV. There
is an underlying continuum in the high excitation-energy region in
the spectrum; it is reasonable to assume that this continuum,
remaining after elimination of the instrumental
background, is primarily due to contributions from excitation of the higher multipoles and quasifree knockout processes \cite{brand}.

\section{Data Analysis}
We have employed the MDA procedure~\cite{exp-bb-1} to extract the strengths of
the isoscalar giant monopole resonance (ISGMR), the isoscalar giant dipole
resonance (ISGDR), the isoscalar giant quadrupole resonance (ISGQR), and 
the high-energy octupole resonance (HEOR) in the Sn isotopes. The
cross-section data were binned into 1-MeV energy intervals to reduce
the statistical fluctuations. For each excitation energy bin from
8.5 MeV to 31.5 MeV, the experimental 17-point angular distribution
$\frac{d\sigma^{exp}}{d\Omega}(\theta_{cm},E_x)$ has been fitted by
means of the least-square method with the linear combination of
calculated distributions
$\frac{d\sigma^{cal}_{L}}{d\Omega}(\theta_{cm},E_x)$ defined by:
\begin{equation}
\frac{d\sigma^{exp}}{d\Omega}(\theta_{cm},E_x)=\sum_{L=0}^{7}a_L(E_x)
\times \frac{d\sigma^{cal}_{L}}{d\Omega}(\theta_{cm},E_x)
\end{equation}
\noindent where  $\frac{d\sigma^{cal}_{L}}{d\Omega}(\theta_{cm},E_x)$ is
the calculated distorted-wave Born approximation (DWBA) cross
section corresponding to 100\% energy-weighted sum rule (EWSR) for
the L-th multipole. The fractions of the EWSR, $a_L(E_x)$, for
various multipole components were determined by minimizing $\chi^2$.
This procedure is justified since the angular distributions are well
characterized by the transferred angular momentum $L$, according to
the DWBA calculations for $\alpha$ scattering. It was confirmed that the
MDA fits were not affected by including $L>$7.

The DWBA calculations were performed following the method of
Satchler and Khoa \cite{th-GRS-1}, using the density-dependent single-folding model
for the real part, obtained with a Gaussian $\alpha$-nucleon potential, 
and a phenomenological Woods-Saxon potential for the imaginary term.
Therefore, the $\alpha$-nucleus interaction is given by:

\begin{equation}
\label{potential}
U(r)=V_F(r)+iW/(1+exp((r-R_I)/a_I)) 
\end{equation}
\noindent where $V_F(R)$ is the real single-folding potential obtained by folding the ground-state density with the density-dependent $\alpha$-nucleon interaction:
\begin{equation}
v_{DDG}(r,r',\rho)=-v(1-\beta\rho(r')^{2/3})exp(-|r-r'|^2/t^2)) \nonumber
\end{equation}

\noindent where $v_{DDG}(r,r',\rho)$ is the density-dependent $\alpha$-nucleon
interaction, $|r-r'|$ is the distance between the center of mass of the $\alpha$
particle and a target nucleon, $\rho(r')$ is the ground-state density of the
target nucleus at the position $r'$ of the target nucleon, $\beta$=1.9 fm$^2$,
and $t$=1.88 fm. W is the depth of the Woods-Saxon type imaginary
part of the potential, with the reduced radius $R_I$ and diffuseness $a_I$.

These calculations were performed with the computer code
PTOLEMY~\cite{ptolemy,ptolemy2}, with the input values
modified~\cite{satchler} to take into account the correct
relativistic kinematics. The shape of the real part of the potential
and the form factor for PTOLEMY were obtained using the codes
SDOLFIN and DOLFIN \cite{dolfin}. We use the transition
densities and sum rules for various multipolarities described in
Refs.~\cite{mnh1,satchler2,mnh3}. The radial moments were
obtained by numerical integration of the Fermi mass distribution
with the parameter values from Ref.~\cite{data-1} (listed in Table~\ref{tab:fermi}).

\begin{table}
\caption{\label{tab:fermi}Fermi-distribution parameters from Ref.~\cite{data-1}. ``c'' is the adjusted half-density radius for the charge distribution and ``a'' is the diffuseness parameter.}
\begin{ruledtabular}
\begin{tabular}{cccccccc}
Target&$^{112}$Sn&$^{114}$Sn&$^{116}$Sn&$^{118}$Sn&$^{120}$Sn&$^{122}$Sn&$^{124}$Sn \\\hline
 c$ (fm)$&5.3714&5.3943&5.4173&5.4391&5.4588&5.4761&5.4907 \\
 a$ (fm)$&0.523&0.523&0.523&0.523&0.523&0.523&0.523 \\
\end{tabular}
\end{ruledtabular}
\end{table}

\begin{figure}[htpb]
\includegraphics[scale=0.8]{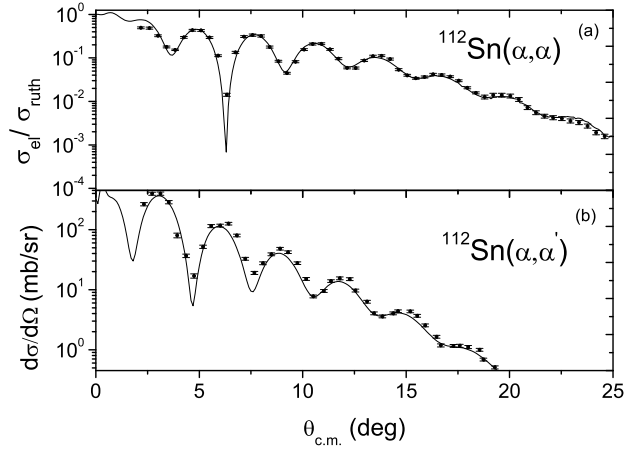}% Here is how to import EPS art
\caption{\label{fig:112sn-elas} (a) Ratio of the elastic $\alpha$-scattering cross sections to the
Rutherford cross sections for $^{112}$Sn at 386 MeV. (b) Differential cross sections for excitation
of the 2$^+_1$ state in $^{112}$Sn. The
solid lines are the results of the folding-model calculations.}
\end{figure}

\begin{figure}[htpb]
\includegraphics[scale=0.8]{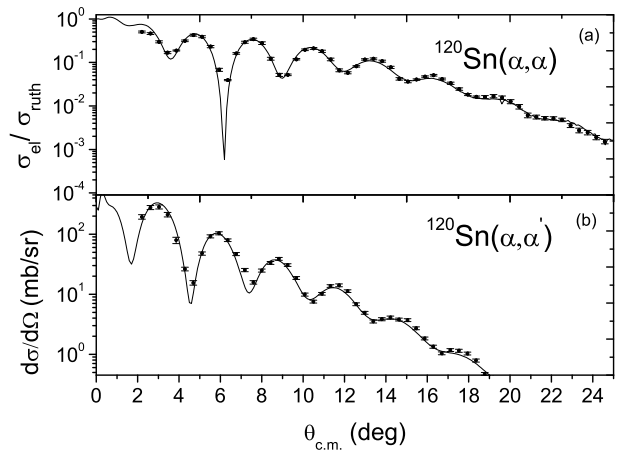}% Here is how to import EPS art
\caption{\label{fig:120sn-elas} Same as Fig.~\ref{fig:112sn-elas}, except for
$^{120}$Sn.}
\end{figure}

\begin{figure}[htpb]
\includegraphics[scale=0.8]{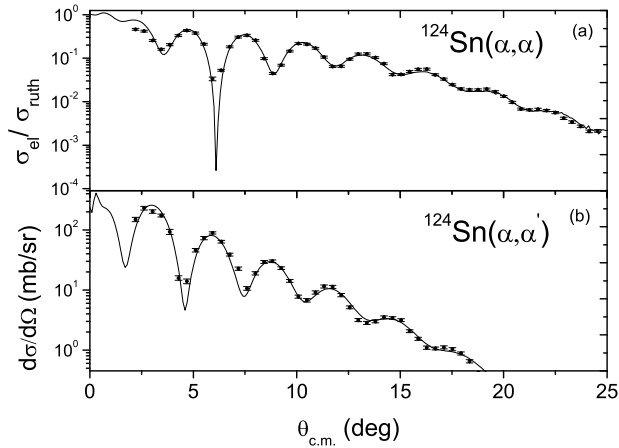}% Here is how to import EPS art
\caption{\label{fig:124sn-elas} Same as Fig.~\ref{fig:112sn-elas}, except for
$^{124}$Sn.}
\end{figure}

The optical-model (OM) parameters {\em viz.} the real part of the potential ($V_F(r)$),
the Woods-Saxon type imaginary part of potential ($W$), the reduced radius ($R_I$), and
the diffuseness ($a_I$) in Eq.~\ref{potential} were determined by fitting
the differential cross sections of elastic $\alpha$ scattering measured for $^{112}$Sn, $^{120}$Sn,
and $^{124}$Sn in a companion experiment; the results are listed in Table~\ref{tab:potential}. 
The OM fits to the elastic scattering data for $^{112}$Sn,
$^{120}$Sn, and $^{124}$Sn, are shown in Figs.~\ref{fig:112sn-elas}(a),~\ref{fig:120sn-elas}(a), 
and \ref{fig:124sn-elas}(a), respectively. 
To test the efficacy of the OM parameters, DWBA calculations were carried out
for the first 2$^+$ states in these nuclei
using a collective form factor and previously-established B(E2) values obtained from 
Refs.~\cite{bel1,bel2} (also listed in Table~\ref{tab:potential}). Figures.~\ref{fig:112sn-elas}(b),
~\ref{fig:120sn-elas}(b), and~\ref{fig:124sn-elas}(b) compare
the results of these calculations with the experimental data; indeed, the DWBA
calculations reproduce the experimental differential cross sections for the 2$^+_1$
states well without any normalization.

\begin{table}
\caption{\label{tab:potential}OM parameters obtained by fitting elastic scattering data.
Also listed are the B(E2) values for the corresponding 2$^+_1$ states from 
Refs.~\cite{bel1,bel2}.}
\begin{ruledtabular}
\begin{tabular}{ccccccc}
Target&V (MeV)&W (MeV)&$a_I$ (fm)&$R_I$ (fm)& B(E2) ($e^2b^2$) \\\hline
$^{112}$Sn&33.9 & 31.7 & 0.60 & 1.02 &0.24\\
$^{120}$Sn&33.4 & 33.0 & 0.63 & 1.01 & 0.20\\
$^{124}$Sn&34.0 & 33.5 & 0.61 & 1.02 & 0.17\\
\end{tabular}
\end{ruledtabular}
\end{table}

The contribution of the IVGDR excitation to the measured cross
sections was subtracted prior to multipole decomposition. Cross
sections for exciting the IVGDR were obtained with DWBA calculations on the basis of the
Goldhaber-Teller model and using the strength
distribution obtained from photonuclear work \cite{Dietrich}. 

\begin{figure}[htpb]
\includegraphics[scale=1]{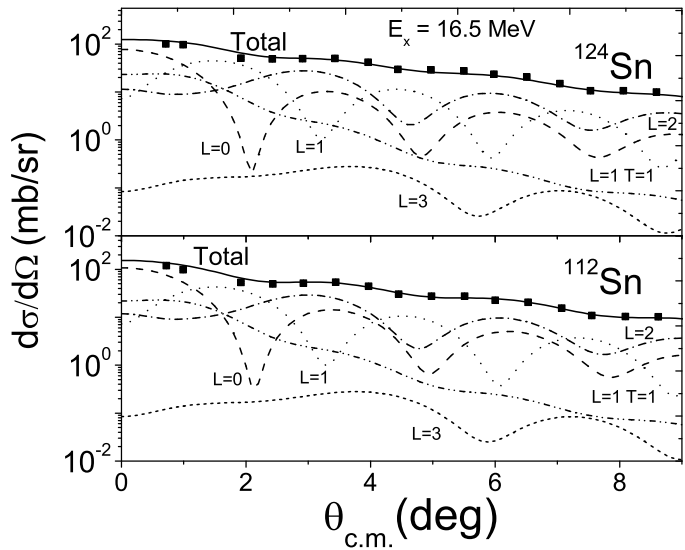}
\caption{\label{fig:ang} Angular distribution of 1-MeV bins centered
at E$_x$=16.5 MeV for $^{112}$Sn($\alpha$,$\alpha'$) and
$^{124}$Sn($\alpha$,$\alpha'$). The solid squares are the
experimental data and the solid lines are the MDA fits to the data.
Also shown are the contributions to the fits from $L$=0 (dashed line),
$L$=1 (dotted line), $L$=2 (dash-dotted line) and $L$=3 (small-dashed line)
multipoles, as well as from the IVGDR (dash-dot-dotted line).}
\end{figure}

\begin{figure}[htpb]
\includegraphics[scale=1]{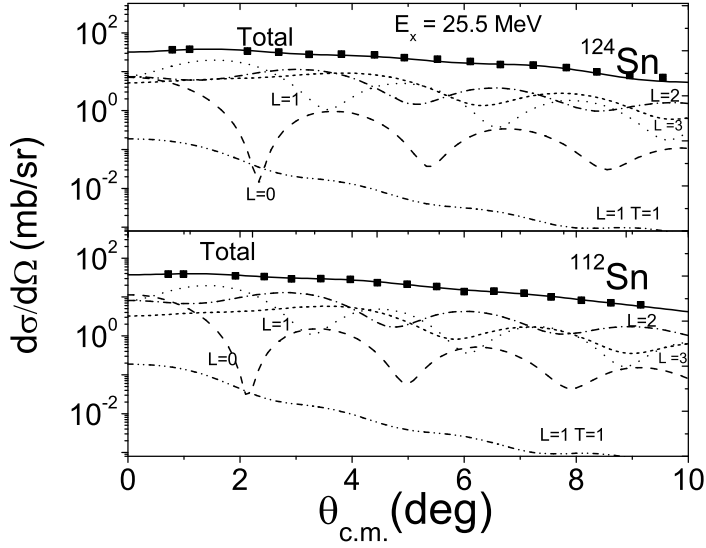}
\caption{\label{fig:ang2} 
Same as Fig.~\ref{fig:ang} except for the 25.5-MeV energy bin (see text).}
\end{figure}

Figs.~\ref{fig:ang} and~\ref{fig:ang2} show the MDA fits to the experimental
angular distributions of the differential cross sections
for the 16.5-MeV and 25.5-MeV energy bins in the inelastic-scattering spectra of
$^{112}$Sn and $^{124}$Sn, respectively, along with the
contributions from the $L$=0, 1 and 2 multipoles. The ISGMR contribution is dominant in
comparison to the other multipoles at E$_x$=16.5 MeV. On the other hand, 
the ISGDR is the dominant contributor at E$_x$=25.5 MeV.

\section{Results and Discussion}

\begin{figure}
\includegraphics[scale=1.4]{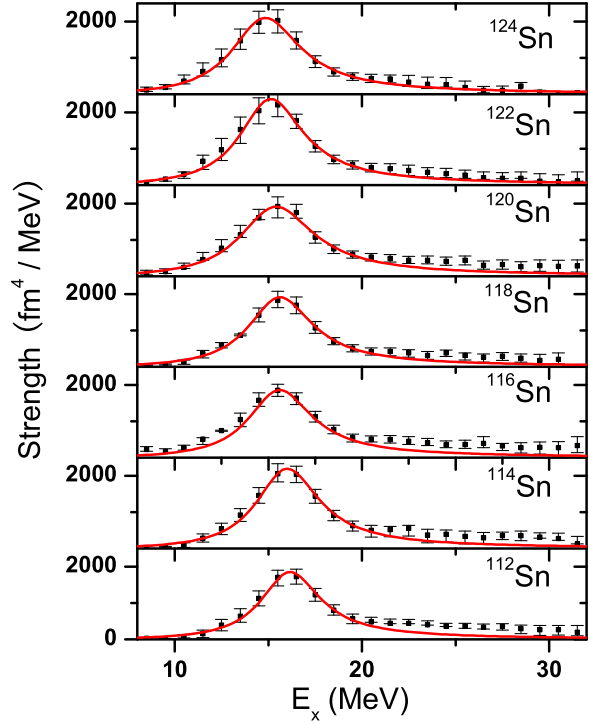}
\caption{\label{fig:ISGMR} (Color online) ISGMR strength distributions obtained for the
Sn isotopes in the present experiment. Error bars represent the
uncertainties from fitting the angular distributions in the MDA procedure.
The solid lines show Lorentzian fits to the data.}
\end{figure}

\begin{figure}
\includegraphics[scale=1.4]{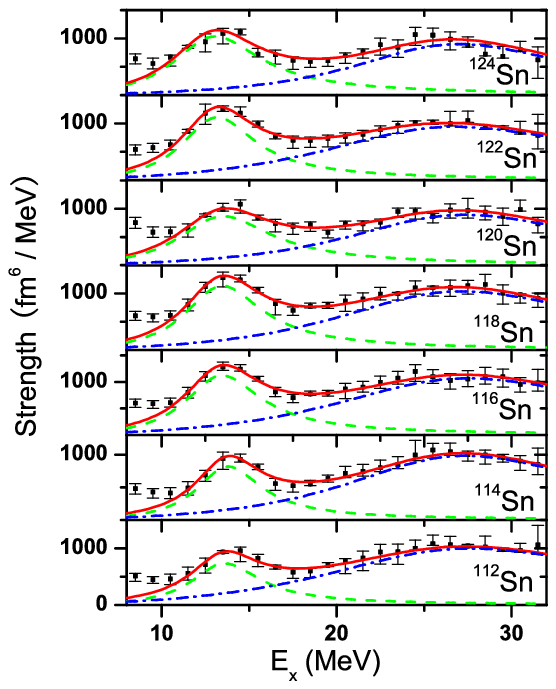}
\caption{\label{fig:gdr} (Color online) ISGDR strength distributions obtained for the
Sn isotopes in the present experiment. Error bars represent the
uncertainties from fitting the angular distributions in the MDA procedure.
The solid lines show Lorentzian fits to the data.}
\end{figure}

\begin{figure}
\includegraphics[scale=1.4]{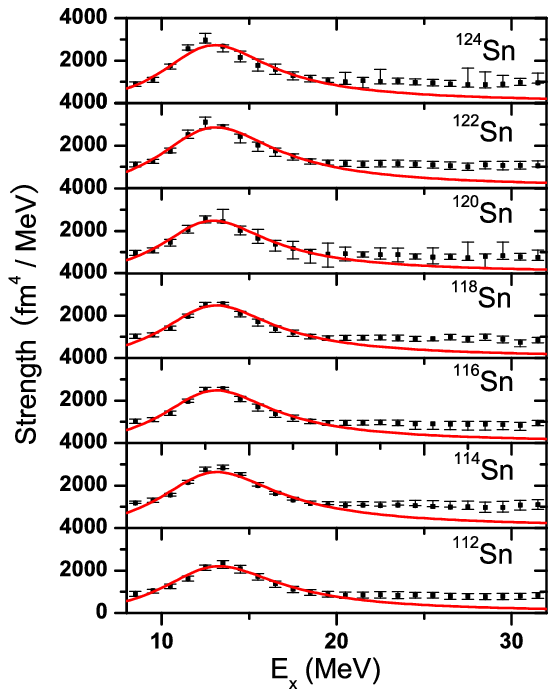}
\caption{\label{fig:ISGQR} (Color online) ISGQR strength distributions obtained for the
Sn isotopes in the present experiment. Error bars represent the
uncertainties from fitting the angular distributions in the MDA procedure.
The solid lines show Lorentzian fits to the data.}
\end{figure}

\begin{figure}[htbp]
\center
\includegraphics[scale=1.2]{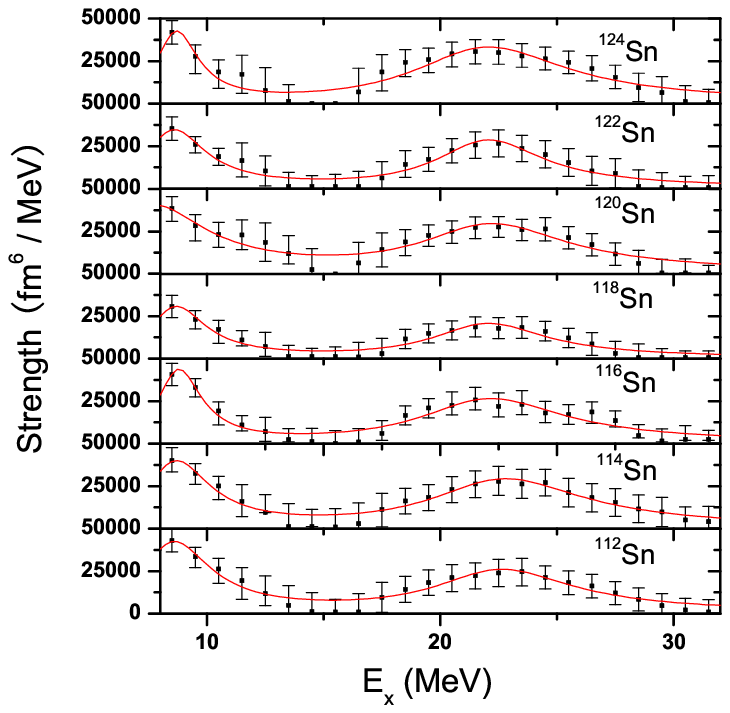}
\caption{\label{fig:heor} (Color online) HEOR strength distributions obtained for
the Sn isotopes in the present experiment. Error bars represent the
uncertainties from fitting the angular distributions in the MDA procedure.
The solid lines show Lorentzian fits to the data.}
\end{figure}

We have extracted strength distributions for $L$=0, 1, 2, and 3
multipoles over the energy range 8.5 MeV--31.5 MeV in all the Sn isotopes
investigated in this work. These are displayed in Figs.~\ref{fig:ISGMR},
\ref{fig:gdr},~\ref{fig:ISGQR}, and~\ref{fig:heor}, respectively.
The strengths are related to the coefficients $a_L$ in Eq.~1
as (see, Refs. ~\cite{mnh1,mnh3}):

\begin{equation}
S_0(E_x)=\frac{2\hbar^2A<r^2>}{mE_x}a_0(E_x)
\end{equation}

\begin{eqnarray}
S_1(E_x)=\frac{3\hbar^2A}{32\pi mE_x}(11<r^4>-\frac{25}{3}<r^2>^2
%\nonumber \\
-10\epsilon<r^2>)a_1(E_x)
\end{eqnarray}

\begin{equation}
S_{L\geq2}(E_x)=\frac{\hbar^2A}{8\pi
mE_x}L(2L+1)^2<r^{2L-2}>a_2(E_x)
\end{equation}

\noindent where $m$, $A$ and $<r^N>$ are the nucleon mass, the mass
number, and the $N$th moment of the ground-state density, and 
$\epsilon$=(4/$E_2$+5/$E_0$)$\hbar^2$/3$mA$; $E_0$ and $E_2$ are
the centroid energies of the ISGMR and the ISGQR and have been taken as
80~A$^{-1/3}$ MeV and 64~A$^{-1/3}$ MeV, respectively.

It should be noted that although we employed calculated DWBA cross sections with up to $L$=7 in the
MDA fitting procedure, it was not possible to reliably extract the strength distributions for $L\geq$4 
because of the limited angular range (0$^{\circ}$--9$^{\circ}$). Further, there is a small,
near-constant ISGMR and ISGQR strength up to the highest excitation energies
measured in this experiment. The {\em raison d`\^{e}tre} of this extra strength is not quite well understood. However, similarly
enhanced E1 strengths at high excitation energies were noted previously
\cite{exp-RCNP-5,exp-RCNP-1} and have been attributed to contributions to the
continuum from three-body channels, such as knockout reactions \cite{brand}.
These processes are 
implicitly included in the MDA as background and may lead to spurious contributions to the
extracted multipole strengths at higher energies where the associated cross sections are very small.
This conjecture is supported by measurements of proton decay from the ISGDR 
at backward angles wherein no
such spurious strength is observed in spectra in coincidence with the decay
protons~\cite{exp-ND-3,hun1,nayak2,hun3}; quasifree knockout results in protons that are forward peaked. A similar
increase in the ISGMR strength at high excitation energies was reported as well by the TAMU group
in $^{12}$C when they carried out MDA without subtracting the continuum from the
excitation-energy spectra~\cite{john}.

The $L=0$ strength distributions were fitted with a Lorentzian
function to determine the centroid energies and widths of the ISGMR. These
fits are shown superimposed in Fig.~\ref{fig:ISGMR}; the corresponding fitting parameters are presented 
in Table~\ref{tbl:ISGMRfit} and compared with results from TAMU~\cite{exp-TAM-6, exp-TAM-3}. In
this and subsequent comparisons and discussion, we refer only to the recent TAMU results because
those are from comparable data and analysis---all other previous results on the Sn isotopes
were from peak-fitting analyses of data taken at significantly lower energies.

In order to compare with the available theoretical results, various
moment ratios for the experimental ISGMR strength distributions have been
calculated over the excitation-energy range, $E_x$ = 10.5--20.5 MeV, 
encompassing the ISGMR peak. The results are listed in
Table~\ref{tbl:gmrmom}.
The reasons for the difference between the present results and those from TAMU for $^{112}$Sn and
$^{124}$Sn are not readily apparent but
might be attributable to the fact that in their analysis the
multipole decomposition is carried out after subtracting a
``background'' from the excitation-energy spectrum, whereas, as
pointed out earlier, no such subtraction is required in the present
analysis since the Sn($\alpha,\alpha'$) spectra obtained in our 
work have been rendered free of all instrumental background events.

Figure~\ref{fig:gdr} shows the strength distributions of ISGDR.
We observe a ``bi-modal'' distribution between $E_x$=8.5 MeV and
$E_x$=31.5 MeV. This bi-modal pattern for the ISGDR has been observed in all
nuclei investigated so far, both in the RCNP and TAMU measurements.
This ``low-energy'' isoscalar $L$=1 strength (LE) has engendered considerable
interest and argument over the past few years. It is present in nearly
all of the recent theoretical calculations in some form or the
other, and at similar energies, although with varying strength. It
has been shown \cite{co00,dv00} that the centroid of this component
of the $L$=1 strength is independent of the nuclear incompressibility
and while the exact nature of this component is not fully
understood yet, suggestions have been extended to the effect that this component might
represent the ``toroidal''~\cite{balb,dv00} or the ``vortex'' modes~\cite{dubna}. 
It is impossible to 
distinguish between the competing possibilities based on currently-available
data~\cite{exp-ND-3}. There is general agreement, however, that only the
high-energy (HE) component of this bi-modal distribution needs to be
considered in obtaining a value of $K_\infty$ from the energy of the ISGDR.
The strength distributions of the ISGDR, therefore, have been fitted with a two-Lorentzian
function and the fitting parameters for the LE- and HE-components are presented
in Tables~\ref{tbl:gdrfitLE} and~\ref{tbl:gdrfitHE}, respectively. It may
be noted that because of the ``spurious'' strength at the higher excitation
energies mentioned previously, the numbers for the extracted EWSR are
significantly larger than 100\% in some cases.

The strength distributions of the ISGQR are shown in Fig.~\ref{fig:ISGQR}. These too were
fitted with a Lorentzian function to determine the centroid energies and the widths.
The fit parameters are
presented in Table~\ref{tbl:ISGQRfit}.

The $L$=3 strength distributions (Fig.~\ref{fig:heor}) show an enhanced
strength at $E_{x} <$ 10 MeV. This part is, most likely, from the
low-energy octupole resonance (LEOR). The LEOR represents the 1$\hbar\omega$
component of the $L$=3 strength and has been reported in a number of nuclei
previously \cite{moss1,moss2}. The strength distributions were, therefore, 
fitted with a two-Lorentzian function to determine the centroid energy of HEOR
(the high-excitation-energy component).
%and, indeed, the peak positions of the lower-energy component for
%$^{116,118,124}$Sn are consistent
%with the values of the LEOR energies reported previously for these
%nuclei \cite{moss1,moss2}.
The extracted HEOR peak-energies are presented in Table~\ref{tbl:heorfit}.

\begin{table}[tpb]
    \caption{\label{tbl:ISGMRfit}Lorentzian-fit parameters for the ISGMR
strength distributions in the Sn isotopes, as extracted from MDA. 
The quoted EWSR values are from the fitted Lorentzians.
The results from TAMU work
(from Gaussian fits), where available, are provided for
comparison~\cite{exp-TAM-6,exp-TAM-3}.}
  \begin{ruledtabular}
    \begin{tabular}{lrrrr}

      \multicolumn{1}{c}{Target} & $E_{ISGMR}$ (MeV)&$\Gamma$ (MeV)&
EWSR\footnotemark[1]& Reference\\
      \hline
      $^{112}$Sn & $16.1\pm0.1$         & $4.0\pm0.4$          & $0.92\pm0.04$        & This work\\
                 & $15.67^{+0.11}_{-0.11}$& $5.18^{+0.40}_{-0.04}$ & $1.10^{+0.15}_{-0.12}$ & TAMU\\

      $^{114}$Sn & $15.9\pm0.1$&$4.1\pm0.4$&$1.04\pm0.06$& This work\\

      $^{116}$Sn & $15.8\pm0.1$&$4.1\pm0.3$&$0.99\pm0.05$& This work\\
       %          & $15.85^{+0.20}_{-0.20}$& $5.27^{+0.25}_{-0.3}$ & $1.12^{+0.15}_{-0.15}$ & TAMU\\

      $^{118}$Sn & $15.6\pm0.1$&$4.3\pm0.4$&$0.95\pm0.05$& This work\\
      $^{120}$Sn & $15.4\pm0.2$&$4.9\pm0.5$&$1.08\pm0.07$& This work\\

      $^{122}$Sn & $15.0\pm0.2$&$4.4\pm0.4$&$1.06\pm0.05$ & This work\\

      $^{124}$Sn & $14.8\pm0.2$&$4.5\pm0.5$&$1.05\pm0.06$ & This work\\
                 & $15.34^{+0.13}_{-0.13}$ & $5.00^{+0.13}_{-0.53}$ & $1.06^{+0.10}_{-0.20}$ & TAMU\\
    \end{tabular}
    \end{ruledtabular}
\footnotetext[1] {
Only statistical uncertainties are included; systematic errors, mostly from
DWBA calculations, are $\sim$15\%.}
\end{table}

\begin{table}[tpb]
    \caption{  \label{tbl:gmrmom}Various
 moment ratios for the ISGMR strength distributions in the Sn
  isotopes.
All moments have been calculated over $E_x$ = 10.5--20.5
MeV. The quoted EWSR values are from the strength observed within this energy
range.
The results from TAMU work, where available,
are provided for comparison~\cite{exp-TAM-6,exp-TAM-3}.}
    \begin{ruledtabular}
    \begin{tabular}{lrrrrr}

      \multicolumn{1}{c}{Target} & $\frac{m_1}{m_0}$ (MeV)&$\sqrt{\frac{m_3}{m_1}}$ (MeV)&$\sqrt{\frac{m_1}{m_{-1}}}$ (MeV)& EWSR\footnotemark[1] &  Reference\\
      \hline
      $^{112}$Sn & $16.2\pm0.1$&$16.7\pm0.2$&$16.1\pm0.1$   & $0.73\pm0.04$        & This work\\
                 & $15.43^{+0.11}_{-0.10}$& $16.05^{+0.26}_{-0.14}$ & $15.23^{+0.10}_{-0.10}$ & $1.16^{+0.13}_{-0.18}$ & TAMU\\

      $^{114}$Sn & $16.1\pm0.1$ & $16.5\pm0.2$ & $15.9\pm0.1$ & $0.86\pm0.05$ & This work\\

      $^{116}$Sn & $15.8\pm0.1$&$16.3\pm0.2$&$15.7\pm0.1$ & $0.86\pm0.05$& This work\\
                 & $15.85\pm0.20$& &    & $1.12\pm0.15$& TAMU\\

      $^{118}$Sn & $15.8\pm0.1$&$16.3\pm0.1$&$15.6\pm0.1$ & $0.73\pm0.04$& This
      work\\

      $^{120}$Sn & $15.7\pm0.1$&$16.2\pm0.2$&$15.5\pm0.1$ & $0.78\pm0.05$& This work\\

      $^{122}$Sn & $15.4\pm0.1$&$15.9\pm0.2$&$15.2\pm0.1$ & $0.85\pm0.05$ & This work\\

      $^{124}$Sn & $15.3\pm0.1$&$15.8\pm0.1$&$15.1\pm0.1$ & $0.77\pm0.05$ & This work\\
                 & $14.50^{+0.14}_{-0.14}$ & $14.96^{+0.10}_{-0.11}$ &$14.33^{+0.17}_{-0.14}$ & $1.04^{+0.11}_{-0.11}$ & TAMU\\
    \end{tabular}
    \end{ruledtabular}
\footnotetext[1] {
Only statistical uncertainties are included; systematic errors, mostly from
DWBA calculations, are $\sim$15\%.}
\end{table}

\begin{table}[tpb]
    \caption{ \label{tbl:gdrfitLE}Lorentzian-fit parameters for the low-energy component of ISGDR strength distributions in the Sn
  isotopes, as extracted from MDA. 
The results from TAMU work, where available,
are provided for comparison~\cite{exp-TAM-6,exp-TAM-3}.}
  \begin{ruledtabular}
    \begin{tabular}{lrrr}

      \multicolumn{1}{c}{Target} & $E_{LE-ISGDR}$ (MeV)&$\Gamma$ (MeV)& Reference\\
      \hline
      $^{112}$Sn & $15.4\pm0.1$         & $4.9\pm0.5$           & This work\\
                 & $14.92^{+0.15}_{-0.14}$& $8.82^{+0.26}_{-0.29}$ & TAMU\\

      $^{114}$Sn & $15.0\pm0.1$&$5.6\pm0.5$& This work\\

      $^{116}$Sn & $14.9\pm0.1$&$5.9\pm0.5$& This work\\
                 & $14.38\pm0.25$& $5.84\pm0.30$ & TAMU\\

      $^{118}$Sn & $14.8\pm0.1$&$6.1\pm0.3$& This work\\

      $^{120}$Sn & $14.7\pm0.1$&$5.9\pm0.3$& This work\\

      $^{122}$Sn & $14.4\pm0.1$&$6.7\pm0.3$& This work\\

      $^{124}$Sn & $14.3\pm0.1$&$6.6\pm0.3$ & This work\\
                 & $13.31^{+0.15}_{-0.15}$ & $6.60^{+0.15}_{-0.13}$  & TAMU\\
    \end{tabular}
\end{ruledtabular}
\end{table}

\begin{table}[tpb]

    \caption{ \label{tbl:gdrfitHE}Lorentzian-fit parameters for the high-energy component of ISGDR strength distributions in the Sn
  isotopes, as extracted from MDA. 
The results from TAMU work, where available,
are provided for comparison~\cite{exp-TAM-6,exp-TAM-3}.}
  \begin{ruledtabular}
    \begin{tabular}{lrrrr}

      \multicolumn{1}{c}{Target} & $E_{HE-ISGDR}$ (MeV)&$\Gamma$ (MeV)
& EWSR\footnotemark[1] &Reference\\
\hline
      $^{112}$Sn & $26.2\pm0.8$         & $16.3\pm4.0$ &$1.02\pm0.03$            & This work\\
                 & $26.28^{+0.32}_{-0.23}$& $10.82^{+0.39}_{-0.36}$ & $0.70^{+0.10}_{-0.10}$& TAMU\\

      $^{114}$Sn & $26.1\pm0.8$&$13.9\pm3.4$& $1.23\pm0.03$& This work\\

      $^{116}$Sn & $25.9\pm0.6$&$13.1\pm4.2$& $1.02\pm0.03$&This work\\
                 & $25.50\pm0.60$& $12.00\pm0.60$ & $0.88\pm0.20$& TAMU\\

      $^{118}$Sn & $26.0\pm0.3$&$13.1\pm2.0$& $1.20\pm0.03$& This work\\

      $^{120}$Sn & $26.0\pm0.4$&$13.1\pm1.9$&$1.50\pm0.03$& This work\\

      $^{122}$Sn & $26.3\pm0.2$&$12.4\pm1.1$ & $1.47\pm0.03$&This work\\

      $^{124}$Sn & $25.7\pm0.5$&$10.2\pm1.6$ & $1.29\pm0.06$&This work\\
                 & $25.06^{+0.22}_{-0.21}$ & $13.87^{+0.28}_{-0.22}$ & $0.93^{+0.12}_{-0.13}$& TAMU\\
    \end{tabular}
\end{ruledtabular}
\footnotetext[1] {
Only statistical uncertainties are included; systematic errors, mostly from
DWBA calculations and the contributions at the highest energies (see text), are $\sim$30\%.}

\end{table}

\begin{table}[tpb]
    \caption{  \label{tbl:ISGQRfit}Lorentzian-fit parameters of ISGQR strength distributions in the Sn
  isotopes, as extracted from MDA. 
The results from TAMU work, where available, are provided for
comparison~\cite{exp-TAM-6,exp-TAM-3}.}
  \begin{ruledtabular}
    \begin{tabular}{lrrrr}

      \multicolumn{1}{c}{Target} & $E_{ISGQR}$ (MeV)&$\Gamma$ (MeV)& 
EWSR\footnotemark[1]&Reference\\
      \hline
      $^{112}$Sn & $13.4\pm0.1$         & $7.0\pm0.5$ &$1.08\pm0.04$          & This work\\
                 & $13.48^{+0.15}_{-0.14}$& $4.90^{+0.22}_{-0.27}$ &$0.88^{+0.14}_{-0.13}$& TAMU\\

      $^{114}$Sn & $13.2\pm0.1$&$6.8\pm0.4$& $1.25\pm0.05$ & This work\\

      $^{116}$Sn & $13.1\pm0.1$&$6.4\pm0.4$&  $1.12\pm0.04$ &This work\\
                 & $13.50\pm0.35$& $5.00\pm0.30$ & $1.08\pm0.12$& TAMU\\

      $^{118}$Sn & $13.1\pm0.1$&$6.6\pm0.3$& $1.08\pm0.03$& This work\\

      $^{120}$Sn & $12.9\pm0.1$&$7.0\pm0.7$& $1.04\pm0.04$& This work\\

      $^{122}$Sn & $12.8\pm0.1$&$7.8\pm0.6$& $1.25\pm0.04$& This work\\

      $^{124}$Sn & $12.6\pm0.1$&$7.7\pm0.9$& $1.13\pm0.04$& This work\\
                 & $12.72^{+0.11}_{-0.11}$ & $4.20^{+0.32}_{-0.03}$ &
$0.89^{+0.15}_{-0.10}$& TAMU\\
    \end{tabular}
\end{ruledtabular}
\footnotetext[1] {
Only statistical uncertainties are included; systematic errors, mostly from
DWBA calculations, are $\sim$20\%.}

\end{table}

\begin{table}[tpb]

    \caption{  \label{tbl:heorfit}Lorentzian-fit parameters of HEOR strength distributions in the Sn
  isotopes, as extracted from MDA. 
The results from TAMU work, where available, are provided for
comparison~\cite{exp-TAM-6,exp-TAM-3}.}
  \begin{ruledtabular}
    \begin{tabular}{lrrr}

      \multicolumn{1}{c}{Target} & $E_{HEOR}$ (MeV)&$\Gamma$ (MeV)& 
Reference\\
      \hline
      $^{112}$Sn & $22.7\pm0.7$         & $7.2\pm1.9$          & This work\\
                 & $20.63^{+0.30}_{-0.28}$& $3.21^{+0.30}_{-0.28}$& TAMU\footnotemark[1]\\

      $^{114}$Sn & $22.7\pm0.7$&$7.2\pm2.1$& This work\\

      $^{116}$Sn & $22.3\pm0.6$&$7.6\pm1.7$& This work\\
                 & $23.3\pm0.8$& $10.9\pm0.6$ & TAMU\\

      $^{118}$Sn & $22.1\pm0.6$&$5.9\pm1.5$& This work\\

      $^{120}$Sn & $22.3\pm0.6$&$7.5\pm1.8$& This work\\

      $^{122}$Sn & $22.1\pm0.6$&$5.6\pm1.5$& This work\\

      $^{124}$Sn & $22.1\pm0.5$&$8.1\pm1.5$& This work\\
                 & $19.12^{+0.26}_{-0.26}$& $3.30^{+0.17}_{-0.05}$& TAMU\footnotemark[1]\\
    \end{tabular}

\end{ruledtabular}
\footnotetext[1] {($m_1/m_0$) ratios.}
\end{table}

\begin{figure}[htbp]
\includegraphics[scale=0.4]{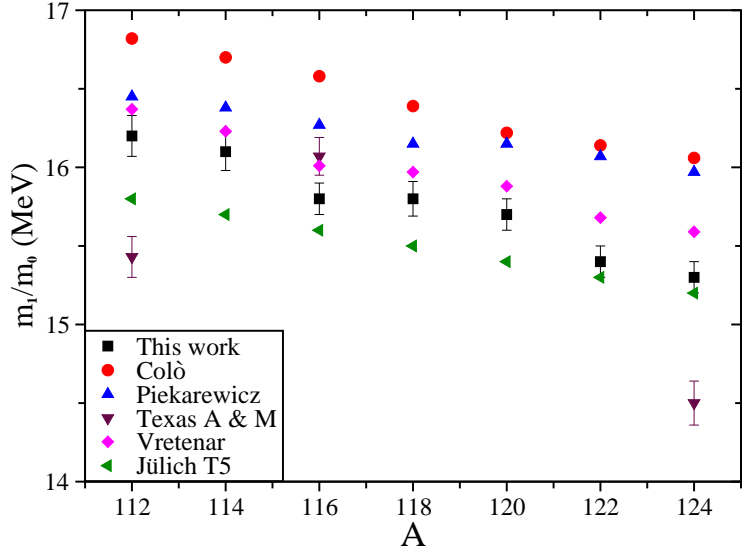}
%\vspace*{-1.0cm}
\caption{\label{fig:snall} (Color online) Systematics of the moment ratios $m_1/m_0$
for the ISGMR strength distributions in the Sn isotopes. The experimental
results (filled squares) are compared with results
from non-relativistic RPA calculations (without pairing) by
Col\`o {\em et al.}~\cite{th-GC-3,th-GC-4} (filled circles);
relativistic calculations of Piekarewicz~\cite{th-JP-4} (triangles); RMF
calculations from Vretenar {\em et al.}~\cite{dario2} (diamonds); and, QTBA
calculations from the J\"{u}lich group~\cite{julich} (sideways triangles).
Results for $^{112}$Sn, $^{116}$Sn and $^{124}$Sn reported by the TAMU
group~\cite{exp-TAM-6,exp-TAM-3} are also shown (inverted triangles). 
}
\end{figure}

The primary focus of this work has been on the ISGMR because of its direct connection
with the nuclear incompressibility. The excitation energy of the ISGMR is expressed in
the scaling model \cite{th-SS-4} as:
\begin{equation}
\label{eqn:gmr} E_{ISGMR}=\hbar\sqrt{\frac{K_A}{m<r^2>}}
\end{equation}

\noindent where $m$ is the nucleon mass, $<r^2>$ the ground-state
mean-square radius, and $K_{A}$, the incompressibility of the
nucleus. 

The moment ratios, $m_1/m_0$, for the ISGMR strengths in the Sn
isotopes are shown in Fig.~\ref{fig:snall} and compared with recent
theoretical results from Col\`o {\em et al.} (non-relativistic)
\cite{th-GC-3,th-GC-4} and Piekarewicz
(relativistic)~\cite{th-JP-4}. The
calculations overestimate the experimental ISGMR energies
significantly (by almost 1 MeV in case of the higher-A isotopes!).
This difference is very surprising since the interactions used in these
calculations are those that very closely reproduce the ISGMR energies
in $^{208}$Pb and $^{90}$Zr. Indeed, this disagreement leaves open a puzzling
question: Why are the tin isotopes so soft \cite{th-JP-4}?
Are there any nuclear structure effects that need to be taken into
account to describe the ISGMR energies in the Sn isotopes? Or, more
provocatively, do the ISGMR energies depend on something more than the
nuclear incompressibility, requiring a modification of the scaling
relationship given in Eq.~\ref{eqn:gmr}?

There have been several attempts to explain this anomaly. 
One of the earliest was by Civitarese {\em et al.}~\cite{civi} to estimate the
effect of pairing correlations on the energy of the ISGMR. The shifts obtained
for the ISGMR energies of 100--150 keV across the Sn isotopic chain were 
insufficient to explain the experimental data.
Piekarewicz and 
Centelles \cite{jorge4} have constructed a hybrid model having the same
incompressibility coefficient ($K_{\infty}$=230 MeV) as the FSUGold \cite{th-JP-3}
while preserving the stiff symmetry energy of NL3 \cite{lala}. This results in a
considerably softer incompressibility coefficient for neutron-rich matter and
produces a significant improvement in agreement with the experimental data on
the ISGMR's in the Sn isotopes. However, as the authors point out, while the improvement in case
of the Sn isotopes is unquestionable, an important problem remains: the hybrid
model underestimates the ISGMR centroid energy in $^{208}$Pb by almost 1 MeV,
suggesting that the rapid softening with neutron excess predicted by this hybrid
model might be unrealistic. They also suggest that the failure of the
FSUGold to reproduce the ISGMR energies might be due to missing physics unrelated to the incompressibility of neutron-rich nuclear matter; as an example of such missing physics, they 
mention the superfluid character of the Sn isotopes resulting from their open-shell structure.

Calculations have also become available recently from the RMF approach with the DD-ME2 interaction \cite{dario1}, and these reproduce the centroids of the ISGMR in the Sn isotopes rather well \cite{dario2}. It is also seen that the DD-ME2 interaction falls within the constraints imposed by the experimental 
$K_{\infty}$ and $K_{\tau}$ values (see discussion below). Some concern has been expressed, however, that this agreement of the centroid energies might be just a coincidence since the ISGMR strength distributions for the Sn isotopes from this work appear to be not significantly different from those obtained from, for example, the FSUGold~\cite{jorgen}.

In calculations using the T5 Skyrme interaction within the quasiparticle time blocking approximation (QTBA) approach, Tselyaev {\em et al.} \cite{julich} have obtained the ISGMR strength distributions in all the Sn isotopes in good agreement with the experimental data, including the resonance widths. However, T5 has the associated $K_{\infty}$ value of only 202 MeV, which is significantly lower than that extracted earlier from the ISGMR's in $^{208}$Pb and $^{90}$Zr. While the agreement with the experimental data is impressive (and, indeed, reproduces the A-dependence rather well), it does leave the question of ``softness'' of the Sn nuclei unanswered. As the authors themselves state, the goal of their work has not been to solve the problem of the nuclear-matter incompressibility but to find under which conditions one can obtain reasonable description of the experimental data for the considered tin isotopes.

The ``superfluid'' character of the Sn isotopes, resulting from pairing
correlations in open-shell nuclei, has been investigated by
Li {\em et al.} \cite{colo3}. In a self-consistent QRPA model that employs the
canonical HFB basis and an
energy-density functional with a Skyrme mean-field part and density-dependent
pairing, they calculated the energy of the ISGMR for the Sn isotopes and looked
at the effects of different kinds of pairing forces (volume, surface, and
mixed). They find that, compared with the HF+RPA and HF-BCS-QRPA formalisms, the
HFB+QRPA calculations lead to energies for the ISGMR in Sn isotopes that are
significantly closer to the experimental values, in particular with the 
surface pairing forces
and the SKM* interaction ($K_{\infty}\sim$ 215 MeV) \cite{skm}. Thus, while
pairing effects lower the ISGMR excitation energies, one still needs to reduce
the $K_{\infty}$ value by $\sim$10\% for achieving a reasonable agreement with
the experimental data.

A very intriguing possibility in explaining the ``softness'' of the Sn isotopes has been offered very recently by Khan \cite{khan,khan2}. The author asserts that, in analogy with the mutually-enhanced-magicity (MEM) effect observed in predictions
of masses with different energy-density functionals \cite{mem1,mem2}, the ISGMR energy in the doubly-magic nuclei might be anomalously higher. The obvious
implication is that the calculations using interactions that are successful in
describing the ISGMR
in the doubly-magic nucleus $^{208}$Pb would necessarily overestimate the ISGMR
energies in the open-shell nuclei. If this effect is manifested in any
significant way, the energy of the ISGMR in the non-doubly-magic Pb isotopes,
$^{204}$Pb and $^{206}$Pb, would be measurably lower than that in
$^{208}$Pb~\cite{khan2}. In the only measurement of the
ISGMR in $^{206}$Pb reported so far~\cite{mnh4}, this conjecture does not appear to hold. Still, precise
measurements of the ISGMR in the Pb isotopes, using background-free inelastic spectra with high-energy
$\alpha$ beams, would be worthwhile to fully examine this possibility.

The incompressibility of a nucleus, $K_{A}$, may be expressed as:
\begin{equation}
K_{A} \sim  K_{vol}(1 + cA^{-1/3}) + K_{\tau}((N - Z)/A)^{2} +
K_{Coul}Z^{2}A^{-4/3}
\end{equation}

\noindent Here, c $\approx$ -1\cite{skp02}, and $K_{Coul}$ is essentially
model independent (in the sense that the deviations from one
theoretical model to another are quite small), so that the
associated term can be calculated for a given isotope. Thus, for a
series of isotopes, the difference $K_A-K_{Coul}Z^2A^{-4/3}$ may be
approximated to have a quadratic relationship with the asymmetry
parameter, of the type y = A + Bx$^2$, with K$_\tau$ being the
coefficient, B, of the quadratic term. It 
has been established previously~\cite{shlomo2,pears} that direct fits to the
Eq.~8 do not provide good constraints on the value of
K$_\infty$. However, this expression is being used here not to
obtain a value for K$_\infty$, but, rather, only to demonstrate the
approximately quadratic relationship between K$_A$ and the asymmetry
parameter.

From such an analysis of the ISGMR data in the Sn isotopes, we have obtained a value of
$K_\tau = -550\pm100$~MeV (see Fig. 4 in Ref.~\cite{tao}). This number
is consistent with the value of $K_{\tau}=-370\pm120$~MeV obtained from
an analysis of the isotopic transport ratios in medium-energy heavy-ion
reactions \cite{bao3}. Incidentally, this value has been modified from the
value of $-500\pm50$~MeV that was quoted previously by this
group~\cite{exp-BAL-1,exp-BAL-4} and referred to in Ref.~\cite{tao}. It
transpires that they had identified the quantity that they had obtained,
$K_{asy}$, as being identical to $K_{\tau}$, the quantity that has been obtained
from the ISGMR measurements; the two differ by a higher-order term
\cite{sagawa6,bao3}. More recently, a value of 
$K_{\tau}=-500^{+125}_{-100}$~MeV
has been obtained by  Centelles {\em et al.} \cite{spain} from constraints put
by neutron-skin data from anti-protonic atoms across the mass
table; here again, it would appear that what the authors have termed
$K_{\tau}$ is actually the aforementioned $K_{asy}$.
Further, a value of $K_{\tau}=-500\pm50$ MeV has been obtained also by
Sagawa {\em et al.} by comparing our Sn ISGMR data with calculations using
different Skyrme Hamiltonians and RMF Lagrangians \cite{sagawa7}. The $K_{\tau}$
value obtained from our ISGMR measurements has, thus, been verified by a number of different procedures involving quite different data.
A more precise determination of K$_\tau$ will likely result
from extending the ISGMR measurements to longer isotopic chains. This provides
strong motivation for measuring the ISGMR strength
in unstable nuclei, a focus of current investigations at the new rare isotope beam facilities
at RIKEN, GANIL, GSI, and NSCL~\cite{hb07,cm07,gsi,nscl}.

Combined with the value of $K_\infty = 240\pm10$~MeV obtained from the ISGMR and ISGDR
data~\cite{exp-ND-3,th-GC-3,th-JP-3,th-SS-2}, we now have ``experimental''
values of both K$_\infty$
and K$_\tau$ which, together, can provide a means of selecting the
most appropriate of the interactions used in EOS calculations. For
example, this combination of ``experimental'' values for K$_\infty$ and K$_\tau$
essentially rules out a vast majority of the Skyrme-type
interactions currently in use in nuclear structure calculations~\cite{sagawa6,ugbulk}. 
Similar conclusions were reached for EOS equations in Refs.~\cite{exp-BAL-3,cs}.

\section{Summary}
We have measured the strength distributions of the isoscalar giant
resonances (ISGMR, ISGDR, ISGQR, and HEOR) in the even-A
$^{112-124}$Sn isotopes via inelastic scattering of 386-MeV $\alpha$
particles at extremely forward angles, including 0$^{\circ}$.  The extracted
parameters for these resonances are in good agreement with previously-obtained
values where available.
% for $^{112}$Sn, $^{116}$Sn, and $^{124}$Sn. 
The ISGMR centroid energies are significantly lower than those 
predicted for these isotopes by recent calculations and point to the need for
further theoretical exploration
of applicable nuclear structure effects, especially the role of pairing in ISGMR strength
calculations in the open-shell nuclei. The asymmetry-term,
$K_{\tau}$, in the expression for the nuclear incompressibility has
been determined to be $-550 \pm 100$ MeV from the ISGMR data in Sn isotopes and 
is found to be consistent with a number of indirectly extracted values for this parameter.

\section{Acknowledgments}
We wish to thank the RCNP staff for providing high-quality $\alpha$
beams required for these measurements. This work has been supported
in part by the US-Japan Cooperative Science Program of the JSPS, and
by the National Science Foundation (Grants No. INT03-42942, PHY04-57120, and PHY07-58100).

\bibliography{snprcrev}%

\begin{thebibliography}{100}
\expandafter\ifx\csname natexlab\endcsname\relax\def\natexlab#1{#1}\fi
\expandafter\ifx\csname bibnamefont\endcsname\relax
  \def\bibnamefont#1{#1}\fi
\expandafter\ifx\csname bibfnamefont\endcsname\relax
  \def\bibfnamefont#1{#1}\fi
\expandafter\ifx\csname citenamefont\endcsname\relax
  \def\citenamefont#1{#1}\fi
\expandafter\ifx\csname url\endcsname\relax
  \def\url#1{\texttt{#1}}\fi
\expandafter\ifx\csname urlprefix\endcsname\relax\def\urlprefix{URL }\fi
\providecommand{\bibinfo}[2]{#2}
\providecommand{\eprint}[2][]{\url{#2}}

\bibitem[{\citenamefont{Pitthan and Walcher}(1971)}]{rp1}
\bibinfo{author}{\bibfnamefont{R.}~\bibnamefont{Pitthan}} \bibnamefont{and}
  \bibinfo{author}{\bibfnamefont{T.}~\bibnamefont{Walcher}},
  \bibinfo{journal}{Phys. Lett.} \textbf{\bibinfo{volume}{36B}},
  \bibinfo{pages}{563} (\bibinfo{year}{1971}).

\bibitem[{\citenamefont{Fukuda and Torizuka}(1972)}]{futo}
\bibinfo{author}{\bibfnamefont{S.}~\bibnamefont{Fukuda}} \bibnamefont{and}
  \bibinfo{author}{\bibfnamefont{Y.}~\bibnamefont{Torizuka}},
  \bibinfo{journal}{Phys.\ Rev.\ Lett.} \textbf{\bibinfo{volume}{29}},
  \bibinfo{pages}{1109} (\bibinfo{year}{1972}).

\bibitem[{\citenamefont{Lewis and Bertrand}(1972)}]{lebe}
\bibinfo{author}{\bibfnamefont{M.~B.} \bibnamefont{Lewis}} \bibnamefont{and}
  \bibinfo{author}{\bibfnamefont{F.~E.} \bibnamefont{Bertrand}},
  \bibinfo{journal}{Nucl.\ Phys.\ A} \textbf{\bibinfo{volume}{196}},
  \bibinfo{pages}{337} (\bibinfo{year}{1972}).

\bibitem[{\citenamefont{{Harakeh {\em et al.}}}(1977)}]{mnh2}
\bibinfo{author}{\bibfnamefont{M.~N.} \bibnamefont{{Harakeh {\em et al.}}}},
  \bibinfo{journal}{Phys. Rev. Lett.} \textbf{\bibinfo{volume}{38}},
  \bibinfo{pages}{686} (\bibinfo{year}{1977}).

\bibitem[{\citenamefont{{Youngblood {\em et al.}}}(1977)}]{dhy2}
\bibinfo{author}{\bibfnamefont{D.~H.} \bibnamefont{{Youngblood {\em et al.}}}},
  \bibinfo{journal}{Phys. Rev. Lett.} \textbf{\bibinfo{volume}{39}},
  \bibinfo{pages}{1188} (\bibinfo{year}{1977}).

\bibitem[{\citenamefont{van~der Woude}(1991)}]{av1}
\bibinfo{author}{\bibfnamefont{A.}~\bibnamefont{van~der Woude}},
  \bibinfo{journal}{Int. Rev. Nucl. Phys.} \textbf{\bibinfo{volume}{7}},
  \bibinfo{pages}{100} (\bibinfo{year}{1991}).

\bibitem[{\citenamefont{Youngblood et~al.}(1981)\citenamefont{Youngblood,
  Bogucki, Bronson, Garg, Lui, and Rozsa}}]{dhy1}
\bibinfo{author}{\bibfnamefont{D.~H.} \bibnamefont{Youngblood}},
  \bibinfo{author}{\bibfnamefont{P.}~\bibnamefont{Bogucki}},
  \bibinfo{author}{\bibfnamefont{J.~D.} \bibnamefont{Bronson}},
  \bibinfo{author}{\bibfnamefont{U.}~\bibnamefont{Garg}},
  \bibinfo{author}{\bibfnamefont{Y.-W.} \bibnamefont{Lui}}, \bibnamefont{and}
  \bibinfo{author}{\bibfnamefont{C.~M.} \bibnamefont{Rozsa}},
  \bibinfo{journal}{Phys. Rev. C} \textbf{\bibinfo{volume}{23}},
  \bibinfo{pages}{1997} (\bibinfo{year}{1981}).

\bibitem[{\citenamefont{{Sharma {\em et al.}}}(1988)}]{sharma}
\bibinfo{author}{\bibfnamefont{M.~M.} \bibnamefont{{Sharma {\em et al.}}}},
  \bibinfo{journal}{Phys.\ Rev. C} \textbf{\bibinfo{volume}{38}},
  \bibinfo{pages}{2562} (\bibinfo{year}{1988}).

\bibitem[{\citenamefont{Morsch et~al.}(1980)\citenamefont{Morsch, Rogge, Turek,
  and Mayer-Boricke}}]{hpm1}
\bibinfo{author}{\bibfnamefont{H.~P.} \bibnamefont{Morsch}},
  \bibinfo{author}{\bibfnamefont{M.}~\bibnamefont{Rogge}},
  \bibinfo{author}{\bibfnamefont{P.}~\bibnamefont{Turek}}, \bibnamefont{and}
  \bibinfo{author}{\bibfnamefont{C.}~\bibnamefont{Mayer-Boricke}},
  \bibinfo{journal}{Phys. Rev. Lett.} \textbf{\bibinfo{volume}{45}},
  \bibinfo{pages}{337} (\bibinfo{year}{1980}).

\bibitem[{\citenamefont{{Davis {\em et al.}}}(1997)}]{bfd1}
\bibinfo{author}{\bibfnamefont{B.~F.} \bibnamefont{{Davis {\em et al.}}}},
  \bibinfo{journal}{Phys. Rev. Lett.} \textbf{\bibinfo{volume}{79}},
  \bibinfo{pages}{607} (\bibinfo{year}{1997}).

\bibitem[{\citenamefont{{Blaizot {\em et al.}}}(1995)}]{th-JPB-1}
\bibinfo{author}{\bibfnamefont{J.}~\bibnamefont{{Blaizot {\em et al.}}}},
  \bibinfo{journal}{Nucl.\ Phys.\ A} \textbf{\bibinfo{volume}{591}},
  \bibinfo{pages}{435} (\bibinfo{year}{1995}).

\bibitem[{\citenamefont{Speth}(1991)}]{speth}
\bibinfo{author}{\bibfnamefont{J.}~\bibnamefont{Speth}},
  \emph{\bibinfo{title}{Electric and Magnetic Giant Resonances in Nuclei}}
  (\bibinfo{publisher}{World Scientific, Singapore}, \bibinfo{year}{1991}).

\bibitem[{\citenamefont{Harakeh and van~der Woude}(2001)}]{mnh1}
\bibinfo{author}{\bibfnamefont{M.~N.} \bibnamefont{Harakeh}} \bibnamefont{and}
  \bibinfo{author}{\bibfnamefont{A.}~\bibnamefont{van~der Woude}},
  \emph{\bibinfo{title}{Giant Resonances: Fundamental High-Frequency Modes of
  Nuclear Excitation}} (\bibinfo{publisher}{Oxford Univ. Press, New York},
  \bibinfo{year}{2001}).

\bibitem[{\citenamefont{{Youngblood {\em et
  al.}}}(1999{\natexlab{a}})}]{exp-TAM-8}
\bibinfo{author}{\bibfnamefont{D.~H.} \bibnamefont{{Youngblood {\em et al.}}}},
  \bibinfo{journal}{Phys. Rev. Lett.} \textbf{\bibinfo{volume}{82}},
  \bibinfo{pages}{691} (\bibinfo{year}{1999}{\natexlab{a}}).

\bibitem[{\citenamefont{{Youngblood {\em et
  al.}}}(1999{\natexlab{b}})}]{exp-TAM-9}
\bibinfo{author}{\bibfnamefont{D.~H.} \bibnamefont{{Youngblood {\em et al.}}}},
  \bibinfo{journal}{Phys.\ Rev.\ C} \textbf{\bibinfo{volume}{60}},
  \bibinfo{pages}{014304} (\bibinfo{year}{1999}{\natexlab{b}}).

\bibitem[{\citenamefont{Clark et~al.}(2001)\citenamefont{Clark, Lui, and
  Youngblood}}]{exp-TAM-7}
\bibinfo{author}{\bibfnamefont{H.~L.} \bibnamefont{Clark}},
  \bibinfo{author}{\bibfnamefont{Y.-W.} \bibnamefont{Lui}}, \bibnamefont{and}
  \bibinfo{author}{\bibfnamefont{D.~H.} \bibnamefont{Youngblood}},
  \bibinfo{journal}{Phys.\ Rev.\ C} \textbf{\bibinfo{volume}{63}},
  \bibinfo{pages}{031301(R)} (\bibinfo{year}{2001}).

\bibitem[{\citenamefont{Youngblood et~al.}(2002)\citenamefont{Youngblood, Lui,
  and Clark}}]{exp-TAM-1}
\bibinfo{author}{\bibfnamefont{D.~H.} \bibnamefont{Youngblood}},
  \bibinfo{author}{\bibfnamefont{Y.-W.} \bibnamefont{Lui}}, \bibnamefont{and}
  \bibinfo{author}{\bibfnamefont{H.~L.} \bibnamefont{Clark}},
  \bibinfo{journal}{Phys.\ Rev.\ C} \textbf{\bibinfo{volume}{65}},
  \bibinfo{pages}{034302} (\bibinfo{year}{2002}).

\bibitem[{\citenamefont{John et~al.}(2003)\citenamefont{John, Tokimoto, Lui,
  Clark, and Youngblood}}]{john}
\bibinfo{author}{\bibfnamefont{B.}~\bibnamefont{John}},
  \bibinfo{author}{\bibfnamefont{Y.}~\bibnamefont{Tokimoto}},
  \bibinfo{author}{\bibfnamefont{Y.-W.} \bibnamefont{Lui}},
  \bibinfo{author}{\bibfnamefont{H.}~\bibnamefont{Clark}}, \bibnamefont{and}
  \bibinfo{author}{\bibfnamefont{D.}~\bibnamefont{Youngblood}},
  \bibinfo{journal}{Phys.\ Rev.\ C} \textbf{\bibinfo{volume}{68}},
  \bibinfo{pages}{014305} (\bibinfo{year}{2003}).

\bibitem[{\citenamefont{{Youngblood {\em et
  al.}}}(2004{\natexlab{a}})}]{exp-TAM-3}
\bibinfo{author}{\bibfnamefont{D.~H.} \bibnamefont{{Youngblood {\em et al.}}}},
  \bibinfo{journal}{Phys.\ Rev.\ C} \textbf{\bibinfo{volume}{69}},
  \bibinfo{pages}{034315} (\bibinfo{year}{2004}{\natexlab{a}}).

\bibitem[{\citenamefont{{Youngblood {\em et
  al.}}}(2004{\natexlab{b}})}]{exp-TAM-2}
\bibinfo{author}{\bibfnamefont{D.~H.} \bibnamefont{{Youngblood {\em et al.}}}},
  \bibinfo{journal}{Phys.\ Rev.\ C} \textbf{\bibinfo{volume}{69}},
  \bibinfo{pages}{054312} (\bibinfo{year}{2004}{\natexlab{b}}).

\bibitem[{\citenamefont{{Lui {\em et al.}}}(2004)}]{exp-TAM-6}
\bibinfo{author}{\bibfnamefont{Y.-W.} \bibnamefont{{Lui {\em et al.}}}},
  \bibinfo{journal}{Phys.\ Rev.\ C} \textbf{\bibinfo{volume}{70}},
  \bibinfo{pages}{014307} (\bibinfo{year}{2004}).

\bibitem[{\citenamefont{{Lui {\em et al.}}}(2006)}]{exp-TAM-5}
\bibinfo{author}{\bibfnamefont{Y.-W.} \bibnamefont{{Lui {\em et al.}}}},
  \bibinfo{journal}{Phys.\ Rev.\ C} \textbf{\bibinfo{volume}{73}},
  \bibinfo{pages}{014314} (\bibinfo{year}{2006}).

\bibitem[{\citenamefont{{Tokimoto {\em et al.}}}(2006)}]{exp-TAM-11}
\bibinfo{author}{\bibfnamefont{Y.}~\bibnamefont{{Tokimoto {\em et al.}}}},
  \bibinfo{journal}{Phys.\ Rev.\ C} \textbf{\bibinfo{volume}{74}},
  \bibinfo{pages}{044308} (\bibinfo{year}{2006}).

\bibitem[{\citenamefont{Youngblood et~al.}(2007)\citenamefont{Youngblood, Lui,
  and Clark}}]{exp-TAM-10}
\bibinfo{author}{\bibfnamefont{D.~H.} \bibnamefont{Youngblood}},
  \bibinfo{author}{\bibfnamefont{Y.-W.} \bibnamefont{Lui}}, \bibnamefont{and}
  \bibinfo{author}{\bibfnamefont{H.~L.} \bibnamefont{Clark}},
  \bibinfo{journal}{Phys.\ Rev.\ C} \textbf{\bibinfo{volume}{76}},
  \bibinfo{pages}{027304} (\bibinfo{year}{2007}).

\bibitem[{\citenamefont{{Bonin {\em et al.}}}(1984)}]{exp-bb-1}
\bibinfo{author}{\bibfnamefont{B.}~\bibnamefont{{Bonin {\em et al.}}}},
  \bibinfo{journal}{Nucl.\ Phys.\ A} \textbf{\bibinfo{volume}{430}},
  \bibinfo{pages}{349} (\bibinfo{year}{1984}).

\bibitem[{\citenamefont{{Hedden {\em et al}}}(2002)}]{ugberk}
\bibinfo{author}{\bibfnamefont{M.}~\bibnamefont{{Hedden {\em et al}}}},
  \bibinfo{journal}{AIP Conf. Proc.} \textbf{\bibinfo{volume}{610}},
  \bibinfo{pages}{880} (\bibinfo{year}{2002}).

\bibitem[{\citenamefont{{Itoh {\em et al.}}}(2002)}]{exp-RCNP-3}
\bibinfo{author}{\bibfnamefont{M.}~\bibnamefont{{Itoh {\em et al.}}}},
  \bibinfo{journal}{Phys.\ Lett.\ B} \textbf{\bibinfo{volume}{549}},
  \bibinfo{pages}{58} (\bibinfo{year}{2002}).

\bibitem[{\citenamefont{{Uchida {\em et al.}}}(2003)}]{exp-RCNP-2}
\bibinfo{author}{\bibfnamefont{M.}~\bibnamefont{{Uchida {\em et al.}}}},
  \bibinfo{journal}{Phys.\ Lett.\ B} \textbf{\bibinfo{volume}{557}},
  \bibinfo{pages}{12} (\bibinfo{year}{2003}).

\bibitem[{\citenamefont{{Itoh {\em et al.}}}(2003)}]{exp-RCNP-5}
\bibinfo{author}{\bibfnamefont{M.}~\bibnamefont{{Itoh {\em et al.}}}},
  \bibinfo{journal}{Phys.\ Rev.\ C} \textbf{\bibinfo{volume}{68}},
  \bibinfo{pages}{064602} (\bibinfo{year}{2003}).

\bibitem[{\citenamefont{{Uchida {\em et al.}}}(2004)}]{exp-RCNP-1}
\bibinfo{author}{\bibfnamefont{M.}~\bibnamefont{{Uchida {\em et al.}}}},
  \bibinfo{journal}{Phys.\ Rev.\ C} \textbf{\bibinfo{volume}{69}},
  \bibinfo{pages}{051301(R)} (\bibinfo{year}{2004}).

\bibitem[{\citenamefont{Garg}(2004)}]{exp-ND-3}
\bibinfo{author}{\bibfnamefont{U.}~\bibnamefont{Garg}},
  \bibinfo{journal}{Nucl.\ Phys.\ A} \textbf{\bibinfo{volume}{731}},
  \bibinfo{pages}{3} (\bibinfo{year}{2004}).

\bibitem[{\citenamefont{{Itoh {\em et al.}}}(2004)}]{itoh}
\bibinfo{author}{\bibfnamefont{M.}~\bibnamefont{{Itoh {\em et al.}}}},
  \bibinfo{journal}{Nucl.\ Phys.\ A} \textbf{\bibinfo{volume}{731}},
  \bibinfo{pages}{41} (\bibinfo{year}{2004}).

\bibitem[{\citenamefont{{Nayak {\em et al.}}}(2006)}]{exp-ND-2}
\bibinfo{author}{\bibfnamefont{B.~K.} \bibnamefont{{Nayak {\em et al.}}}},
  \bibinfo{journal}{Phys.\ Lett.\ B} \textbf{\bibinfo{volume}{637}},
  \bibinfo{pages}{43} (\bibinfo{year}{2006}).

\bibitem[{\citenamefont{{Li {\em et al.}}}(2007)}]{tao}
\bibinfo{author}{\bibfnamefont{T.}~\bibnamefont{{Li {\em et al.}}}},
  \bibinfo{journal}{Phys. Rev. Lett.} \textbf{\bibinfo{volume}{99}},
  \bibinfo{pages}{162503} (\bibinfo{year}{2007}).

\bibitem[{\citenamefont{{Garg {\em et al.}}}(2007)}]{ug7}
\bibinfo{author}{\bibfnamefont{U.}~\bibnamefont{{Garg {\em et al.}}}},
  \bibinfo{journal}{Nucl.\ Phys.\ A} \textbf{\bibinfo{volume}{788}},
  \bibinfo{pages}{36} (\bibinfo{year}{2007}).

\bibitem[{\citenamefont{{Fujiwara {\em et al.}}}(1999)}]{exp-RCNP-6}
\bibinfo{author}{\bibfnamefont{M.}~\bibnamefont{{Fujiwara {\em et al.}}}},
  \bibinfo{journal}{Nucl. Instrum. Meth. Phys. Res. A}
  \textbf{\bibinfo{volume}{422}}, \bibinfo{pages}{484} (\bibinfo{year}{1999}).

\bibitem[{\citenamefont{Lui et~al.}(1984)\citenamefont{Lui, Bogucki, Bronson,
  Youngblood, and Garg}}]{dhy4}
\bibinfo{author}{\bibfnamefont{Y.-W.} \bibnamefont{Lui}},
  \bibinfo{author}{\bibfnamefont{P.}~\bibnamefont{Bogucki}},
  \bibinfo{author}{\bibfnamefont{J.~D.} \bibnamefont{Bronson}},
  \bibinfo{author}{\bibfnamefont{D.~H.} \bibnamefont{Youngblood}},
  \bibnamefont{and} \bibinfo{author}{\bibfnamefont{U.}~\bibnamefont{Garg}},
  \bibinfo{journal}{Phys. Rev. C} \textbf{\bibinfo{volume}{30}},
  \bibinfo{pages}{51} (\bibinfo{year}{1984}).

\bibitem[{\citenamefont{{James M.~Lattimer} and {Madappa
  Prakash}}(2000)}]{exp-JML-1}
\bibinfo{author}{\bibnamefont{{James M.~Lattimer}}} \bibnamefont{and}
  \bibinfo{author}{\bibnamefont{{Madappa Prakash}}}, \bibinfo{journal}{Phys.
  Rep.} \textbf{\bibinfo{volume}{333}}, \bibinfo{pages}{121}
  (\bibinfo{year}{2000}).

\bibitem[{\citenamefont{Lattimer and Prakash}(2004)}]{latti}
\bibinfo{author}{\bibfnamefont{J.~M.} \bibnamefont{Lattimer}} \bibnamefont{and}
  \bibinfo{author}{\bibfnamefont{M.}~\bibnamefont{Prakash}},
  \bibinfo{journal}{Science} \textbf{\bibinfo{volume}{304}},
  \bibinfo{pages}{532} (\bibinfo{year}{2004}).

\bibitem[{\citenamefont{{Steiner {\em et al.}}}(2005)}]{stein}
\bibinfo{author}{\bibfnamefont{A.~W.} \bibnamefont{{Steiner {\em et al.}}}},
  \bibinfo{journal}{Phys. Rep.} \textbf{\bibinfo{volume}{411}}
  (\bibinfo{year}{2005}).

\bibitem[{\citenamefont{Li and Steiner}(2006)}]{exp-BAL-2}
\bibinfo{author}{\bibfnamefont{B.-A.} \bibnamefont{Li}} \bibnamefont{and}
  \bibinfo{author}{\bibfnamefont{A.~W.} \bibnamefont{Steiner}},
  \bibinfo{journal}{Phys. Lett. B} \textbf{\bibinfo{volume}{642}},
  \bibinfo{pages}{436} (\bibinfo{year}{2006}).

\bibitem[{\citenamefont{Itoh}()}]{itohthesis}
\bibinfo{author}{\bibfnamefont{M.}~\bibnamefont{Itoh}}, \eprint{Ph.D. thesis,
  Kyoto University (2003)}.

\bibitem[{\citenamefont{{Itoh {\em et al.}}}(1999)}]{itoh-rep}
\bibinfo{author}{\bibfnamefont{M.}~\bibnamefont{{Itoh {\em et al.}}}},
  \bibinfo{journal}{RCNP Annual Report} p.~\bibinfo{pages}{7}
  (\bibinfo{year}{1999}).

\bibitem[{\citenamefont{{Brandenburg {\em et al.}}}(1987)}]{brand}
\bibinfo{author}{\bibfnamefont{S.}~\bibnamefont{{Brandenburg {\em et al.}}}},
  \bibinfo{journal}{Nuc.\ Phys.\ A} \textbf{\bibinfo{volume}{466}},
  \bibinfo{pages}{29} (\bibinfo{year}{1987}).

\bibitem[{\citenamefont{Satchler and Khoa}(1997)}]{th-GRS-1}
\bibinfo{author}{\bibfnamefont{G.~R.} \bibnamefont{Satchler}} \bibnamefont{and}
  \bibinfo{author}{\bibfnamefont{D.~T.} \bibnamefont{Khoa}},
  \bibinfo{journal}{Phys.\ Rev.\ C} \textbf{\bibinfo{volume}{55}},
  \bibinfo{pages}{285} (\bibinfo{year}{1997}).

\bibitem[{\citenamefont{Rhoades-Brown
  et~al.}(1980{\natexlab{a}})\citenamefont{Rhoades-Brown, Macfarlane, and
  Pieper}}]{ptolemy}
\bibinfo{author}{\bibfnamefont{M.}~\bibnamefont{Rhoades-Brown}},
  \bibinfo{author}{\bibfnamefont{M.~H.} \bibnamefont{Macfarlane}},
  \bibnamefont{and} \bibinfo{author}{\bibfnamefont{S.~C.}
  \bibnamefont{Pieper}}, \bibinfo{journal}{Phys. Rev. C}
  \textbf{\bibinfo{volume}{21}}, \bibinfo{pages}{2417}
  (\bibinfo{year}{1980}{\natexlab{a}}).

\bibitem[{\citenamefont{Rhoades-Brown
  et~al.}(1980{\natexlab{b}})\citenamefont{Rhoades-Brown, Macfarlane, and
  Pieper}}]{ptolemy2}
\bibinfo{author}{\bibfnamefont{M.}~\bibnamefont{Rhoades-Brown}},
  \bibinfo{author}{\bibfnamefont{M.~H.} \bibnamefont{Macfarlane}},
  \bibnamefont{and} \bibinfo{author}{\bibfnamefont{S.~C.}
  \bibnamefont{Pieper}}, \bibinfo{journal}{Phys. Rev. C}
  \textbf{\bibinfo{volume}{21}}, \bibinfo{pages}{2436}
  (\bibinfo{year}{1980}{\natexlab{b}}).

\bibitem[{\citenamefont{Satchler}(1992)}]{satchler}
\bibinfo{author}{\bibfnamefont{G.~R.} \bibnamefont{Satchler}},
  \bibinfo{journal}{Nucl. Phys. A} \textbf{\bibinfo{volume}{540}}
  (\bibinfo{year}{1992}).

\bibitem[{\citenamefont{Rickersten}(1976)}]{dolfin}
\bibinfo{author}{\bibfnamefont{L.~D.} \bibnamefont{Rickersten}},
  \bibinfo{journal}{unpublished}  (\bibinfo{year}{1976}).

\bibitem[{\citenamefont{Satchler}(1987)}]{satchler2}
\bibinfo{author}{\bibfnamefont{G.~R.} \bibnamefont{Satchler}},
  \bibinfo{journal}{Nucl. Phys. A} \textbf{\bibinfo{volume}{472}}
  (\bibinfo{year}{1987}).

\bibitem[{\citenamefont{Harakeh and Dieperink}(1981)}]{mnh3}
\bibinfo{author}{\bibfnamefont{M.~N.} \bibnamefont{Harakeh}} \bibnamefont{and}
  \bibinfo{author}{\bibfnamefont{A.~E.~L.} \bibnamefont{Dieperink}},
  \bibinfo{journal}{Phys.\ Rev. C} \textbf{\bibinfo{volume}{23}},
  \bibinfo{pages}{2329} (\bibinfo{year}{1981}).

\bibitem[{\citenamefont{{Frickle {\em et al.}}}(1995)}]{data-1}
\bibinfo{author}{\bibfnamefont{G.}~\bibnamefont{{Frickle {\em et al.}}}},
  \bibinfo{journal}{At.\ Data Nucl.\ Data Tables}
  \textbf{\bibinfo{volume}{60}}, \bibinfo{pages}{2} (\bibinfo{year}{1995}).

\bibitem[{\citenamefont{{Raman {\em et al.}}}(1987)}]{bel1}
\bibinfo{author}{\bibfnamefont{S.}~\bibnamefont{{Raman {\em et al.}}}},
  \bibinfo{journal}{At. Data Nucl. Data Table} \textbf{\bibinfo{volume}{36}}
  (\bibinfo{year}{1987}).

\bibitem[{\citenamefont{Spear}(1989)}]{bel2}
\bibinfo{author}{\bibfnamefont{R.~H.} \bibnamefont{Spear}},
  \bibinfo{journal}{At. Data Nucl. Data Table} \textbf{\bibinfo{volume}{42}}
  (\bibinfo{year}{1989}).

\bibitem[{\citenamefont{Dietrich and Berman}(1988)}]{Dietrich}
\bibinfo{author}{\bibfnamefont{S.~S.} \bibnamefont{Dietrich}} \bibnamefont{and}
  \bibinfo{author}{\bibfnamefont{B.~L.} \bibnamefont{Berman}},
  \bibinfo{journal}{At. Data Nucl. Data Tables} \textbf{\bibinfo{volume}{38}}
  (\bibinfo{year}{1988}).

\bibitem[{\citenamefont{{Hunyadi {\em et al.}}}(2003)}]{hun1}
\bibinfo{author}{\bibfnamefont{M.}~\bibnamefont{{Hunyadi {\em et al.}}}},
  \bibinfo{journal}{Phys.\ Lett.\ B} \textbf{\bibinfo{volume}{576}},
  \bibinfo{pages}{253} (\bibinfo{year}{2003}).

\bibitem[{\citenamefont{{Nayak {\em et al.}}}(2009)}]{nayak2}
\bibinfo{author}{\bibfnamefont{B.~K.} \bibnamefont{{Nayak {\em et al.}}}},
  \bibinfo{journal}{Phys.\ Lett.\ B} \textbf{\bibinfo{volume}{674}},
  \bibinfo{pages}{281} (\bibinfo{year}{2009}).

\bibitem[{\citenamefont{{Hunyadi {\em et al.}}}(2009)}]{hun3}
\bibinfo{author}{\bibfnamefont{M.}~\bibnamefont{{Hunyadi {\em et al.}}}},
  \bibinfo{journal}{Phys.\ Rev. C} \textbf{\bibinfo{volume}{80}},
  \bibinfo{pages}{044317} (\bibinfo{year}{2009}).

\bibitem[{\citenamefont{{Col\`o {\em et al.}}}(2000)}]{co00}
\bibinfo{author}{\bibfnamefont{G.}~\bibnamefont{{Col\`o {\em et al.}}}},
  \bibinfo{journal}{Phys. Lett. B} \textbf{\bibinfo{volume}{485}},
  \bibinfo{pages}{362} (\bibinfo{year}{2000}).

\bibitem[{\citenamefont{Vretenar et~al.}(2000)\citenamefont{Vretenar, Wandelt,
  and Ring}}]{dv00}
\bibinfo{author}{\bibfnamefont{D.}~\bibnamefont{Vretenar}},
  \bibinfo{author}{\bibfnamefont{A.}~\bibnamefont{Wandelt}}, \bibnamefont{and}
  \bibinfo{author}{\bibfnamefont{P.}~\bibnamefont{Ring}},
  \bibinfo{journal}{Phys. Lett. B} \textbf{\bibinfo{volume}{485}},
  \bibinfo{pages}{334} (\bibinfo{year}{2000}).

\bibitem[{\citenamefont{Balbutsev et~al.}(1994)\citenamefont{Balbutsev,
  Molodtsova, and Unzhakova}}]{balb}
\bibinfo{author}{\bibfnamefont{E.~B.} \bibnamefont{Balbutsev}},
  \bibinfo{author}{\bibfnamefont{I.~V.} \bibnamefont{Molodtsova}},
  \bibnamefont{and} \bibinfo{author}{\bibfnamefont{A.~V.}
  \bibnamefont{Unzhakova}}, \bibinfo{journal}{Europhys. Lett.}
  \textbf{\bibinfo{volume}{26}}, \bibinfo{pages}{499} (\bibinfo{year}{1994}).

\bibitem[{\citenamefont{Nesterenko}()}]{dubna}
\bibinfo{author}{\bibfnamefont{V.}~\bibnamefont{Nesterenko}}, \eprint{private
  communication}.

\bibitem[{\citenamefont{{Moss {\em et al.}}}(1976)}]{moss1}
\bibinfo{author}{\bibfnamefont{J.~M.} \bibnamefont{{Moss {\em et al.}}}},
  \bibinfo{journal}{Phys.\ Rev.\ Lett.} \textbf{\bibinfo{volume}{37}},
  \bibinfo{pages}{816} (\bibinfo{year}{1976}).

\bibitem[{\citenamefont{{Moss {\em et al.}}}(1978)}]{moss2}
\bibinfo{author}{\bibfnamefont{J.~M.} \bibnamefont{{Moss {\em et al.}}}},
  \bibinfo{journal}{Phys.\ Rev. C} \textbf{\bibinfo{volume}{18}},
  \bibinfo{pages}{741} (\bibinfo{year}{1978}).

\bibitem[{\citenamefont{{Col\`o {\em et al.}}}(2004)}]{th-GC-3}
\bibinfo{author}{\bibfnamefont{G.}~\bibnamefont{{Col\`o {\em et al.}}}},
  \bibinfo{journal}{Phys.\ Rev.\ C} \textbf{\bibinfo{volume}{70}},
  \bibinfo{pages}{024307} (\bibinfo{year}{2004}).

\bibitem[{\citenamefont{Col\`o}()}]{th-GC-4}
\bibinfo{author}{\bibfnamefont{G.}~\bibnamefont{Col\`o}}, \eprint{private
  communication}.

\bibitem[{\citenamefont{Piekarewicz}(2007)}]{th-JP-4}
\bibinfo{author}{\bibfnamefont{J.}~\bibnamefont{Piekarewicz}},
  \bibinfo{journal}{Phys.\ Rev.\ C} \textbf{\bibinfo{volume}{76}},
  \bibinfo{pages}{031301} (\bibinfo{year}{2007}).

\bibitem[{\citenamefont{Vretenar}()}]{dario2}
\bibinfo{author}{\bibfnamefont{D.}~\bibnamefont{Vretenar}}, \eprint{private
  communication}.

\bibitem[{\citenamefont{{Tselyaev {\em et al.}}}(2009)}]{julich}
\bibinfo{author}{\bibfnamefont{V.}~\bibnamefont{{Tselyaev {\em et al.}}}},
  \bibinfo{journal}{Phys.\ Rev.\ C} \textbf{\bibinfo{volume}{79}},
  \bibinfo{pages}{034309} (\bibinfo{year}{2009}).

\bibitem[{\citenamefont{Stringari}(1982)}]{th-SS-4}
\bibinfo{author}{\bibfnamefont{S.}~\bibnamefont{Stringari}},
  \bibinfo{journal}{Phys.\ Lett. B} \textbf{\bibinfo{volume}{108}},
  \bibinfo{pages}{232} (\bibinfo{year}{1982}).

\bibitem[{\citenamefont{{Civitarese {\em et al.}}}(1991)}]{civi}
\bibinfo{author}{\bibfnamefont{O.}~\bibnamefont{{Civitarese {\em et al.}}}},
  \bibinfo{journal}{Phys.\ Rev. C} \textbf{\bibinfo{volume}{43}},
  \bibinfo{pages}{2622} (\bibinfo{year}{1991}).

\bibitem[{\citenamefont{Piekarewicz and Centelles}(2009)}]{jorge4}
\bibinfo{author}{\bibfnamefont{J.}~\bibnamefont{Piekarewicz}} \bibnamefont{and}
  \bibinfo{author}{\bibfnamefont{M.}~\bibnamefont{Centelles}},
  \bibinfo{journal}{Phys.\ Rev.\ C} \textbf{\bibinfo{volume}{79}},
  \bibinfo{pages}{054311} (\bibinfo{year}{2009}).

\bibitem[{\citenamefont{Todd-Rutel and Piekarewicz}(2005)}]{th-JP-3}
\bibinfo{author}{\bibfnamefont{B.~G.} \bibnamefont{Todd-Rutel}}
  \bibnamefont{and}
  \bibinfo{author}{\bibfnamefont{J.}~\bibnamefont{Piekarewicz}},
  \bibinfo{journal}{Phys.\ Rev.\ Lett.} \textbf{\bibinfo{volume}{95}},
  \bibinfo{pages}{122501} (\bibinfo{year}{2005}).

\bibitem[{\citenamefont{{Lalazissis {\em et al.}}}(1997)}]{lala}
\bibinfo{author}{\bibfnamefont{G.~A.} \bibnamefont{{Lalazissis {\em et al.}}}},
  \bibinfo{journal}{Phys.\ Rev.\ C} \textbf{\bibinfo{volume}{55}},
  \bibinfo{pages}{540} (\bibinfo{year}{1997}).

\bibitem[{\citenamefont{{Lalazissis {\em et al.}}}(2003)}]{dario1}
\bibinfo{author}{\bibfnamefont{G.~A.} \bibnamefont{{Lalazissis {\em et al.}}}},
  \bibinfo{journal}{Phys.\ Rev.\ C} \textbf{\bibinfo{volume}{71}},
  \bibinfo{pages}{024312} (\bibinfo{year}{2003}).

\bibitem[{\citenamefont{Piekarewicz}()}]{jorgen}
\bibinfo{author}{\bibfnamefont{J.}~\bibnamefont{Piekarewicz}}, \eprint{private
  communication}.

\bibitem[{\citenamefont{{Li {\em et al.}}}(2008)}]{colo3}
\bibinfo{author}{\bibfnamefont{J.}~\bibnamefont{{Li {\em et al.}}}},
  \bibinfo{journal}{Phys.\ Rev.\ C} \textbf{\bibinfo{volume}{78}},
  \bibinfo{pages}{064304} (\bibinfo{year}{2008}).

\bibitem[{\citenamefont{{Bartel {\em et al.}}}(1982)}]{skm}
\bibinfo{author}{\bibfnamefont{J.}~\bibnamefont{{Bartel {\em et al.}}}},
  \bibinfo{journal}{Nucl.\ Phys.\ A} \textbf{\bibinfo{volume}{386}},
  \bibinfo{pages}{79} (\bibinfo{year}{1982}).

\bibitem[{\citenamefont{Khan}(2009{\natexlab{a}})}]{khan}
\bibinfo{author}{\bibfnamefont{E.}~\bibnamefont{Khan}},
  \bibinfo{journal}{Phys.\ Rev. C} \textbf{\bibinfo{volume}{80}},
  \bibinfo{pages}{011307} (\bibinfo{year}{2009}{\natexlab{a}}).

\bibitem[{\citenamefont{Khan}(2009{\natexlab{b}})}]{khan2}
\bibinfo{author}{\bibfnamefont{E.}~\bibnamefont{Khan}},
  \bibinfo{journal}{Phys.\ Rev. C} \textbf{\bibinfo{volume}{80}},
  \bibinfo{pages}{057302} (\bibinfo{year}{2009}{\natexlab{b}}).

\bibitem[{\citenamefont{{Lunney {\em et al.}}}(2003)}]{mem1}
\bibinfo{author}{\bibfnamefont{D.}~\bibnamefont{{Lunney {\em et al.}}}},
  \bibinfo{journal}{Rev.\ Mod.\ Phys.} \textbf{\bibinfo{volume}{75}},
  \bibinfo{pages}{1021} (\bibinfo{year}{2003}).

\bibitem[{\citenamefont{{Zeldes {\em et al.}}}(1983)}]{mem2}
\bibinfo{author}{\bibfnamefont{N.}~\bibnamefont{{Zeldes {\em et al.}}}},
  \bibinfo{journal}{Nucl.\ Phys.\ A} \textbf{\bibinfo{volume}{399}},
  \bibinfo{pages}{11} (\bibinfo{year}{1983}).

\bibitem[{\citenamefont{{Harakeh {\em et al.}}}(1979)}]{mnh4}
\bibinfo{author}{\bibfnamefont{M.~N.} \bibnamefont{{Harakeh {\em et al.}}}},
  \bibinfo{journal}{Nucl.\ Phys.\ A} \textbf{\bibinfo{volume}{327}},
  \bibinfo{pages}{373} (\bibinfo{year}{1979}).

\bibitem[{\citenamefont{{Patra {\em et al.}}}(2002)}]{skp02}
\bibinfo{author}{\bibfnamefont{S.}~\bibnamefont{{Patra {\em et al.}}}},
  \bibinfo{journal}{Phys. Rev. C} \textbf{\bibinfo{volume}{65}},
  \bibinfo{pages}{044304} (\bibinfo{year}{2002}).

\bibitem[{\citenamefont{Shlomo and Youngblood}(1993)}]{shlomo2}
\bibinfo{author}{\bibfnamefont{S.}~\bibnamefont{Shlomo}} \bibnamefont{and}
  \bibinfo{author}{\bibfnamefont{D.~H.} \bibnamefont{Youngblood}},
  \bibinfo{journal}{Phys.\ Rev. C} \textbf{\bibinfo{volume}{47}},
  \bibinfo{pages}{529} (\bibinfo{year}{1993}).

\bibitem[{\citenamefont{Pearson}(1991)}]{pears}
\bibinfo{author}{\bibfnamefont{J.~M.} \bibnamefont{Pearson}},
  \bibinfo{journal}{Phys.\ Lett. B} \textbf{\bibinfo{volume}{271}},
  \bibinfo{pages}{12} (\bibinfo{year}{1991}).

\bibitem[{\citenamefont{{Chen {\em et al.}}}(2009)}]{bao3}
\bibinfo{author}{\bibfnamefont{L.-W.} \bibnamefont{{Chen {\em et al.}}}},
  \bibinfo{journal}{Phys.\ Rev. C} \textbf{\bibinfo{volume}{80}},
  \bibinfo{pages}{014322} (\bibinfo{year}{2009}).

\bibitem[{\citenamefont{Li and Chen}(2005)}]{exp-BAL-1}
\bibinfo{author}{\bibfnamefont{B.-A.} \bibnamefont{Li}} \bibnamefont{and}
  \bibinfo{author}{\bibfnamefont{L.-W.} \bibnamefont{Chen}},
  \bibinfo{journal}{Phys. Rev. C} \textbf{\bibinfo{volume}{72}},
  \bibinfo{pages}{064611} (\bibinfo{year}{2005}).

\bibitem[{\citenamefont{{Lie-Wen Chen}
  et~al.}(2005{\natexlab{a}})\citenamefont{{Lie-Wen Chen}, {Che Ming Ko}, and
  {Bao-An Li}}}]{exp-BAL-4}
\bibinfo{author}{\bibnamefont{{Lie-Wen Chen}}},
  \bibinfo{author}{\bibnamefont{{Che Ming Ko}}}, \bibnamefont{and}
  \bibinfo{author}{\bibnamefont{{Bao-An Li}}}, \bibinfo{journal}{Phys. Rev.
  Lett} \textbf{\bibinfo{volume}{94}}, \bibinfo{pages}{032701}
  (\bibinfo{year}{2005}{\natexlab{a}}).

\bibitem[{\citenamefont{Yoshida and Sagawa}(2006)}]{sagawa6}
\bibinfo{author}{\bibfnamefont{S.}~\bibnamefont{Yoshida}} \bibnamefont{and}
  \bibinfo{author}{\bibfnamefont{H.}~\bibnamefont{Sagawa}},
  \bibinfo{journal}{Phys. Rev. C} \textbf{\bibinfo{volume}{73}},
  \bibinfo{pages}{044320} (\bibinfo{year}{2006}).

\bibitem[{\citenamefont{{Centelles {\em et al.}}}(2009)}]{spain}
\bibinfo{author}{\bibfnamefont{M.}~\bibnamefont{{Centelles {\em et al.}}}},
  \bibinfo{journal}{Phys. Rev. Lett.} \textbf{\bibinfo{volume}{102}},
  \bibinfo{pages}{122502} (\bibinfo{year}{2009}).

\bibitem[{\citenamefont{{Sagawa {\em et al.}}}(2007)}]{sagawa7}
\bibinfo{author}{\bibfnamefont{H.}~\bibnamefont{{Sagawa {\em et al.}}}},
  \bibinfo{journal}{Phys. Rev. C} \textbf{\bibinfo{volume}{76}},
  \bibinfo{pages}{034327} (\bibinfo{year}{2007}).

\bibitem[{\citenamefont{{Baba {\em et al.}}}(2007)}]{hb07}
\bibinfo{author}{\bibfnamefont{H.}~\bibnamefont{{Baba {\em et al.}}}},
  \bibinfo{journal}{Nucl. Phys. A} \textbf{\bibinfo{volume}{788}},
  \bibinfo{pages}{188c} (\bibinfo{year}{2007}).

\bibitem[{\citenamefont{{Monrozeau {\em et al.}}}(2008)}]{cm07}
\bibinfo{author}{\bibfnamefont{C.}~\bibnamefont{{Monrozeau {\em et al.}}}},
  \bibinfo{journal}{Phys. Rev. Lett.} \textbf{\bibinfo{volume}{100}},
  \bibinfo{pages}{042501} (\bibinfo{year}{2008}).

\bibitem[{gsi()}]{gsi}
\eprint{http://www.gsi.de/fair/index\_e.html}.

\bibitem[{nsc()}]{nscl}
\eprint{http://www.frib.msu.edu/}.

\bibitem[{\citenamefont{Agrawal et~al.}(2003)\citenamefont{Agrawal, Shlomo, and
  Kim~Au}}]{th-SS-2}
\bibinfo{author}{\bibfnamefont{B.~K.} \bibnamefont{Agrawal}},
  \bibinfo{author}{\bibfnamefont{S.}~\bibnamefont{Shlomo}}, \bibnamefont{and}
  \bibinfo{author}{\bibfnamefont{V.}~\bibnamefont{Kim~Au}},
  \bibinfo{journal}{Phys.\ Rev. C} \textbf{\bibinfo{volume}{68}},
  \bibinfo{pages}{031304} (\bibinfo{year}{2003}).

\bibitem[{\citenamefont{Garg}(2009)}]{ugbulk}
\bibinfo{author}{\bibfnamefont{U.}~\bibnamefont{Garg}}, \bibinfo{journal}{AIP
  Conf. Proc.} \textbf{\bibinfo{volume}{1128}}, \bibinfo{pages}{100}
  (\bibinfo{year}{2009}).

\bibitem[{\citenamefont{{Lie-Wen Chen}
  et~al.}(2005{\natexlab{b}})\citenamefont{{Lie-Wen Chen}, {Che Ming Ko}, and
  {Bao-An Li}}}]{exp-BAL-3}
\bibinfo{author}{\bibnamefont{{Lie-Wen Chen}}},
  \bibinfo{author}{\bibnamefont{{Che Ming Ko}}}, \bibnamefont{and}
  \bibinfo{author}{\bibnamefont{{Bao-An Li}}}, \bibinfo{journal}{Phys. Rev. C}
  \textbf{\bibinfo{volume}{72}}, \bibinfo{pages}{064309}
  (\bibinfo{year}{2005}{\natexlab{b}}).

\bibitem[{\citenamefont{{Basu {\em et al.}}}(2009)}]{cs}
\bibinfo{author}{\bibfnamefont{D.~N.} \bibnamefont{{Basu {\em et al.}}}},
  \bibinfo{journal}{Phys.\ Rev. C} \textbf{\bibinfo{volume}{80}},
  \bibinfo{pages}{057304} (\bibinfo{year}{2009}).

\end{thebibliography}

\end{document}